\documentclass[aps,pre,twocolumn,groupedaddress,showpacs,floatfix]{revtex4}
\pdfoutput=1
\usepackage{dcolumn}
\usepackage{amsmath}
\usepackage{amssymb}
\usepackage{graphicx}
\usepackage{bm}
\usepackage[T1]{fontenc}
\usepackage{color}
\usepackage{url}
\usepackage[bf]{subfigure}
\usepackage{rotating}
\usepackage{scalefnt}
\usepackage{multirow}

\usepackage{epstopdf}

\newcommand{\dSp}{\hbox{$\phi$}}

\newcommand{\pNi}[1] {\hbox{$\bar{i_{#1}}$}}
\newcommand{\pNr}[1] {\hbox{$\bar{r_{#1}}$}}

\newcommand{\Ig}{\hbox{$S$}} 
\newcommand{\Sp}{\hbox{$I$}} 
\newcommand{\St}{\hbox{$R$}}

\usepackage{amsthm}
 
\theoremstyle{definition}

\begin{document}

\title{Rumor propagation with heterogeneous transmission in social networks}

\author{Didier A. Vega-Oliveros}
\affiliation{Instituto de Ci\^{e}ncias Matem\'{a}ticas e de Computa\c{c}\~{a}o, Universidade de S\~{a}o Paulo, CP 668,13560-970 S\~{a}o Carlos, SP, Brazil.}

\author{Luciano da F. Costa}
\affiliation{Instituto de F\'{\i}sica de S\~{a}o Carlos, Universidade de S\~{a}o Paulo, CP 369, 13560-970  S\~{a}o Carlos, SP, Brazil.}

\author{Francisco A. Rodrigues}
\email{francisco@icmc.usp.br}
\affiliation{Departamento de Matem\'{a}tica Aplicada e Estat\'{i}stica, Instituto de Ci\^{e}ncias Matem\'{a}ticas e de Computa\c{c}\~{a}o, Universidade de S\~{a}o Paulo, CP 668, 13560-970 S\~{a}o Carlos, SP, Brazil.}

\begin{abstract}
Rumor models consider that information transmission occurs with the same probability between each pair of nodes. However, this assumption is not observed in social networks, which contain influential spreaders. To overcome this limitation, we assume that central individuals have a higher capacity of convincing their neighbors than peripheral subjects. By extensive numerical simulations we find that the spreading is improved in scale-free networks when the transmission probability is proportional to PageRank, degree, and betweenness centrality. In addition, the results suggest that the spreading can be controlled by adjusting the transmission probabilities of the most central nodes. Our results provide a conceptual framework for understanding the interplay between rumor propagation and heterogeneous transmission in social networks.

\end{abstract}

\pacs{89.75.Hc,89.75.-k,89.75.Kd}

\maketitle

\section{Introduction}

Rumor spreading is one of the most basic mechanisms for information dissemination in society. Some decades ago, rumors circulated only by everyday word-of-mouth conversations. However, today many rumors are either originated or reinforced by electronic media, including social online networks. News, advertising and gossips are propagated quickly in social networks and has an important impact on society, influencing the financial market and causing deep anxiety about epidemic outbreak or terrorist attacks~\cite{Kosfeld05,Kimmel04}. For instance, it has been verified that investors over-react to bad news and reactions to bad news are stronger than those to good news~\cite{Kaminsky99, Schindler07}. 

The theoretical study of rumor dynamics is fundamental to understand phase transitions and develop strategies to control and optimize dissemination~\cite{Pastor-Satorras2001, Hethcote2006, Castellano2009, Funk2010}. Several models have been developed to analyze how information is transmitted in social networks~\cite{Pastor-Satorras2015}. These models consider that the rumor may contaminate individuals as an infectious agent --- their mind is infected with the information. The most traditional model was proposed by Daley and Kendal~\cite{DALEY1964}, which divides a closed and homogeneously mixed population in three compartments, i.e., ignorants, spreaders, and stiflers. The rumor is transmitted through the population by pair-wise contacts between spreaders and ignorants. When a spreader meets another informed individual, both informed subjects may lose the interest in the rumor and do not propagate it anymore. The model proposed by Maki and Thompson~\cite{Maki1973} is similar, but only the initiating spreader can become a stifler after the interaction.

The Daley-Kendal and Maki-Thompson models do not consider the heterogeneous structure of the society~\cite{DALEY1964, Maki1973}. To overcome this limitation, more accurate models have been proposed after the establishment of Network Science~\cite{barabasiEalbert1999, newman2003, boccaletti06}. In this case, most individuals interact with only a few neighbors, whereas a small number of subjects have a huge number of contacts~\cite{Zanette2001a, Moreno2004a, Nekovee2007}. The study of the Daley-Kendal model in networks revealed that the network structure plays a fundamental role in rumor propagation, influencing the reach and velocity of the information spreading~\cite{Pastor-Satorras2001, Moreno2004a}. 

Although the network structure provides a more realistic modeling of information propagation, most models consider homogeneous transmission~\cite{Castellano2009, Pastor-Satorras2015}, i.e., the information is transmitted with the same probability in all contacts. However, this assumption is not observed in social and technological networks. For example, subjects do not propagate information with the same efficiency --- celebrities and political leaders are more influential spreaders than ordinary people. The effect of the heterogeneous propagation, in which the capacity of transmission is not the same for all vertices, has not been addressed in rumor models yet, although it is observed in social networks.

Heterogeneous contagion was treated before in the case of disease spreading. Meloni et al.~\cite{Meloni2009} adopted the susceptible-infected-susceptible (SIS) model in which the capacities for processing and delivering information are distributed according to the betweenness centrality measure. They verified that the value of the epidemic threshold in scale-free networks directly depends on the first and second statistical moments of the betweenness centrality. 

In this present paper, we propose a diffusion model in which the propagation probability is heterogeneously distributed. We assume that central individuals have a higher ability to convince their neighbors than peripheral subjects. This property is observed in social networks, in which densely connected~\cite{Jennings37} or central individuals are very influent~\cite{Kitsak2010, Pei013, Arruda014, Pei014, Morone2015}. The probability of contagion is considered as correlated with centrality measures, including degree, betweenness centrality, closeness centrality, $k$-core, PageRank, and eigenvector centrality. In artificial scale-free networks, the positive correlation between the transmission probability and the degree provides the largest rumor dissemination. On the other hand, the correlation with PageRank and betweenness centrality enables obtaining the highest fraction of informed individuals in social networks. Our simulations results indicate that the correlation between the transmission probability and centrality measures is necessary to enhance the rumor propagation since perturbations in this correlation decrease the reach of the rumor. 

We also provide a method to control rumor dissemination. Our analysis is inspired by the event occurred in the Carnival Magic cruise in Oct., 2014 \footnote{Jack Healy. (2014, October 20). Ebola is ruled out after cruise ship carrying hospital worker returns to Texas. \textit{The New York Times}. Retrieved from http://www.nytimes.com}. A laboratory supervisor had processed clinical samples of a man diagnosed with Ebola. The patient died four days before the ship left. The supervisor was voluntarily quarantined in her cabin to avoid possible infection and spread. A blood sample was taken and the Ebola test results were negative. Authorities assumed there was no evidence of a public health threat to the cruise passengers or any bordering country. However, the anxiety and fear of the passengers, lack of clarity and delayed release of official information by authorities led to panic and misinformation spreading. Thus, to avoid such situation, it is important to develop theoretical approaches to control the rumor propagation. We perform Monte Carlo simulations to reproduce this situation and consider heterogeneous propagation. Our results show that to control the rumor spreading, it is necessary to change the transmission probabilities of the central nodes and act on this changing at the beginning of the rumor diffusion.

The paper is structured as follows. The heterogeneous rumor model is presented in Section~\ref{sec:genModel}. Section~\ref{sec:database} describes artificial and social networks considered in our analysis. Simulation results and applications are provided in Sections~\ref{sec:simulation}. Finally, Section~\ref{sec:final} presents our conclusions. Centrality measures are presented as an appendix at the end of this paper.

\section{Heterogeneous propagation}\label{sec:genModel}

Rumor spreading models assume that the probabilities of propagation (or propagation rates in the continuous time models) are the same for all vertices~\cite{Pastor-Satorras2015}. These models present three compartments: (i) ignorants (\Ig) are unaware of the information, (ii) spreaders or infected (\Sp) disseminate the rumor and (iii) stiflers or removed (\St) know the information, but do not participate in the spreading process anymore. In the Maki-Thompson (MK) rumor model~\cite{Maki1973}, whenever a spreader $x$ contacts an ignorant $y$, the ignorant may become a spreader with a fixed probability $\beta$. Otherwise, if $y$ is informed (\emph{i.e.} a spreader or a stifler) then $x$ may turn into a stifler with probability $\mu$. Thus, the transition to the recovered state does not occur spontaneously as in the epidemic spreading models, but through contacts~\cite{Pastor-Satorras2015}. The possible transitions of the MK model are given by~\cite{Maki1973,Pastor-Satorras2015}
\begin{equation}
\left\{\begin{array}{lcc}
I + S \stackrel{\beta}{\longrightarrow} I + I;\\
I + R \stackrel{\mu}{\longrightarrow}  R + R;\\
I + I \stackrel{\mu}{\longrightarrow} R + I.
\end{array}\right.
\label{eq:modeloDK}
\end{equation} 
where operator $+$ represents the contact between two nodes. The difference with respect to the Daley-Kendall (DK) model~\cite{DALEY1964} lies in the contact between neighboring spreaders. More specifically, in the DK model both spreaders may turn into stiflers when a contact occurs, i.e. $I + I \stackrel{\mu}{\longrightarrow} R + R$.

In addition to the MK and DK models, several other models have been proposed for modeling of rumor spreading on networks, as described in~\cite{Moreno2004a,  Keeling2005, Hethcote2006,Castellano2009,Pastor-Satorras2015}. 

\subsection{The model}\label{sec:CM}

Here, we propose a rumor model based on discrete-time Markov chains with heterogeneous probabilities. The motivation for this model lies in the fact that activity patterns in social dynamics are mostly heterogeneous and not all individuals have the same propensity to convince others (\emph{e.g.}~\cite{Meloni2009}). The probabilities of a node $x$ being ignorant ($\Ig_x(t)$), spreader ($\Sp_x(t)$) or stifler ($\St_x(t)$) satisfies $\Ig_x(t) \: + \: \Sp_x(t) \: + \: \St_x(t) \: = \: 1$ for all time steps $t$. The information spreading can be described by the general rumor rules presented in the set of Equations~\ref{eq:modeloDK}. In this case, the probability of a node $x$ being in one of possible states at time $(t \:+\:1)$ is given by
\begin{equation}
\Ig_x(t+1) = \Ig_x(t) \: \pNi{x}(t) \: , 
\label{eq:hetIg}
\end{equation}
\begin{equation}
\Sp_x(t+1) = \Sp_x(t) \: \pNr{x}(t) + \Ig_x(t)\left[\:1 \: - \: \pNi{x}(t)\:\right] \: ,
\label{eq:hetSp}
\end{equation}
\begin{equation}
\St_x(t+1) = \St_x(t) \; + \; \Sp_x(t)\left[\:1 \: - \: \pNr{x}(t)\:\right] \: ,
\label{eq:hetSt}
\end{equation} 
where $\pNi{x}(t)$ is the probability that vertex $x$ does not become a spreader by any of its neighbors at time $t$. Similarly, $\pNr{x}(t)$ is the probability that $x$ does not become a stifler. In Equation~\ref{eq:hetSp}, vertex $x$ is a spreader at time $(t \:+\:1)$ whether: (i) $x$ was a spreader in the last step (with probability $\Sp_x(t)$) and has not become a stifler during that time, which occurs with probability $\pNr{x}(t)$; or (ii) $x$ was an ignorant at time $t$ (with probability $\Ig_x(t)$) and became informed by any of its neighbors according to the probability $\left[\,1 \, - \, \pNi{x}(t)\,\right]$. The probability of vertex $x$ being a stifler at time $(t \:+\:1)$, Equation~\ref{eq:hetSt}, depends on the following conditions: (i) $x$ was a spreader that became a stifler from any of its informed neighbors, which happens according to probability $\Sp_x(t)\left[\:1 \: - \: \pNr{x}(t)\:\right]$; or (ii) $x$ was a stifler in the last time step (with probability $\St_x(t)$) and remained in this state at time $(t \:+\:1)$. 

The probability that vertex $x$ does not become a spreader or a stifler at time $t$, i.e., the probability that it does not perform a transition to other states is given by
\begin{equation}
 {\pNi{x}}(t) =  \prod^{N}_{y = 1}\left[\: 1 \; - \; \beta_y \: \frac{A_{yx}}{k_y} \; \Sp_y(t)\:\right] \: , 
\label{eq:hetsi}
\end{equation}
\begin{equation}
 {\pNr{x}}(t) =  1 \; - \; \sum^{N}_{y = 1} \: \mu_x \: \frac{A_{xy}}{k_x}\left[\:\Sp_y(t) \;+\; \St_y(t)\:\right]  \: .
\label{eq:hetri}
\end{equation}
The probability of vertex $x$ has not become a spreader depends on whether any of its spreader neighbors has not infected it (Equation \ref{eq:hetsi}). Vertex $x$ may be infected at time $t$ by one of its spreader neighbors $y$ having propagation probability $\beta_y$. Thereby, the probability of $x$ has not became a spreader by $y$ is $\left[ 1 - \beta_y (\frac{A_{yx}}{k_y}) \Sp_y(t)\right]$. The product operator Equation \ref{eq:hetsi} is due to the assumption that each neighbor contacts $x$ independently. In contrast, a spreader $x$ may become recovered by contacting informed neighbors (Equation~\ref{eq:hetri}). If vertex $x$ contacts a neighbor that is informed, then $x$ may become a stifler according to probability $\mu_x$. 

In this formulation, we assume that infection and recovery do not occur in the same time step. In addition, a uniform random contact between a spreader and a neighbor per unit time, known as a contact process (CP), is adopted in our formulation. Thus, the probability to select a neighbor $y$ of $x$ is $A_{xy}/{k_x}$ . 

The heterogeneous propagation is incorporated in the model by introducing a characteristic vector $\vec{\beta}\, = \, \{\beta_1, \: \beta_2, \: \ldots, \: \beta_N \}$, whose elements represent the propagation probability for each vertex. Likewise, vector $\vec{\mu} =  \{\mu_1, \: \mu_2, \: \ldots, \: \mu_N \}$ yields the recovery probabilities. The classical rumor model with homogeneous propagation is recovered by making $\beta_x = \beta$ and $\mu_x = \mu$, $x=1,\ldots, N$, where $\beta$ and $\mu$ are constants.

The fraction of spreaders can be evaluated by iterating Eq. \ref{eq:hetIg}, 
\begin{equation}
S_x(t) = S_x(0)\prod_{u=0}^t  {\pNi{x}}(u).
\end{equation}
Thus, in the long run ($t \rightarrow \infty$) we have $I_x(\infty) = 0$, $x=1,2,\ldots, N$, because the number of stiflers is null. Using $I_x(t) + S_x(t) + R_x(t) = 1$, we calculate the final fraction of stiflers,
\begin{equation}
R_x(\infty) = 1 - S_x(\infty) = 1 - S_x(0)\prod_{u=0}^\infty  {\pNi{x}}(u).
\end{equation}
If the spreading starts in a single seed node, then
\begin{equation}\label{Eq:Rx}
R_x(\infty) = 1 - \frac{1}{N}\prod_{u=0}^\infty \prod^{N}_{y = 1}\left[ 1  - \beta_y  \frac{A_{yx}}{k_y}  \Sp_y(u)\:\right].
\end{equation}

\begin{figure}[!tb]
\centering	  
  \subfigure[]{\label{fig:lambda}\includegraphics[width= 0.5\textwidth]{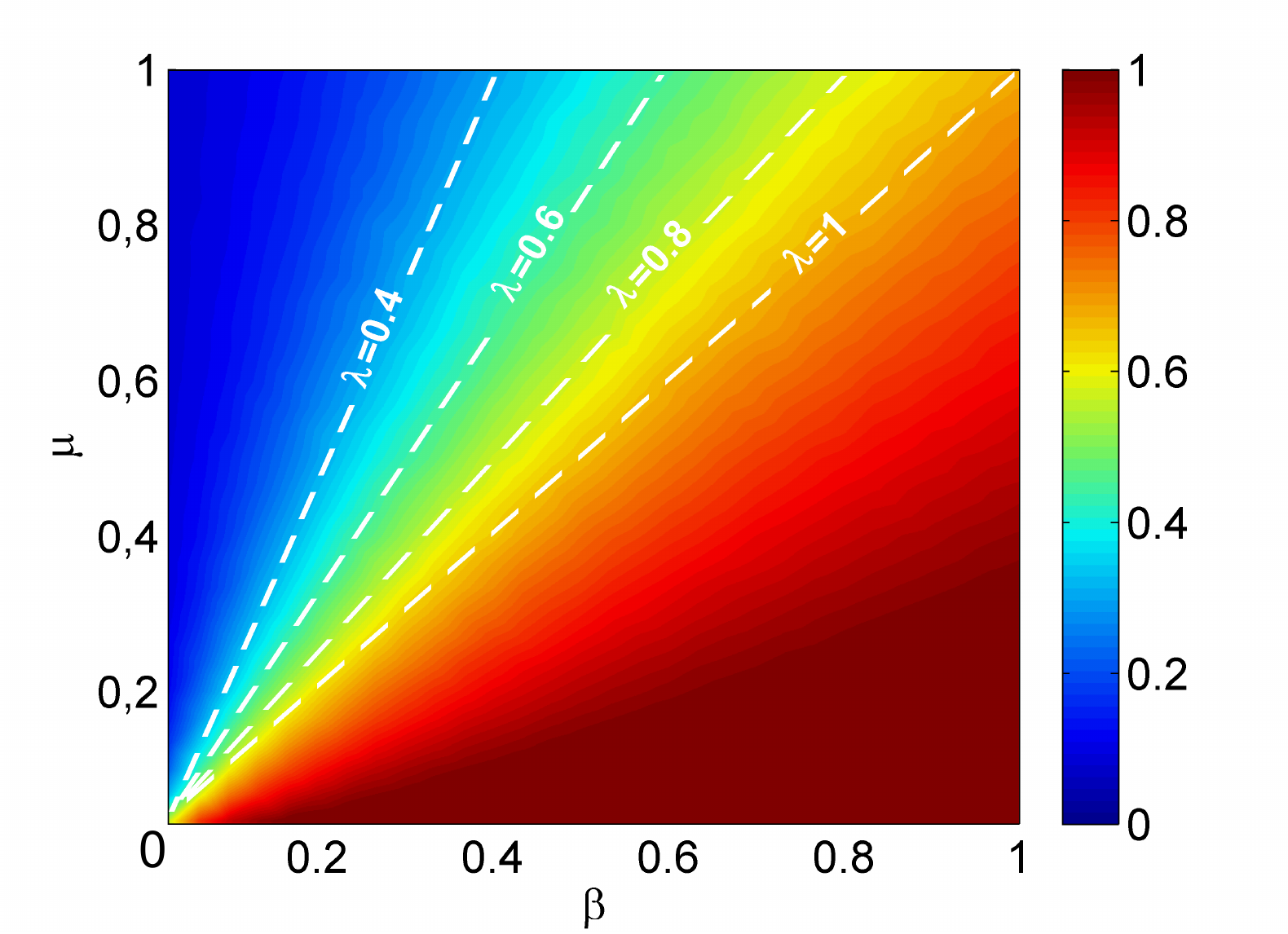}} 
  \subfigure[]{\label{fig:BA10mil50HOM}\includegraphics[width= 0.45\textwidth]{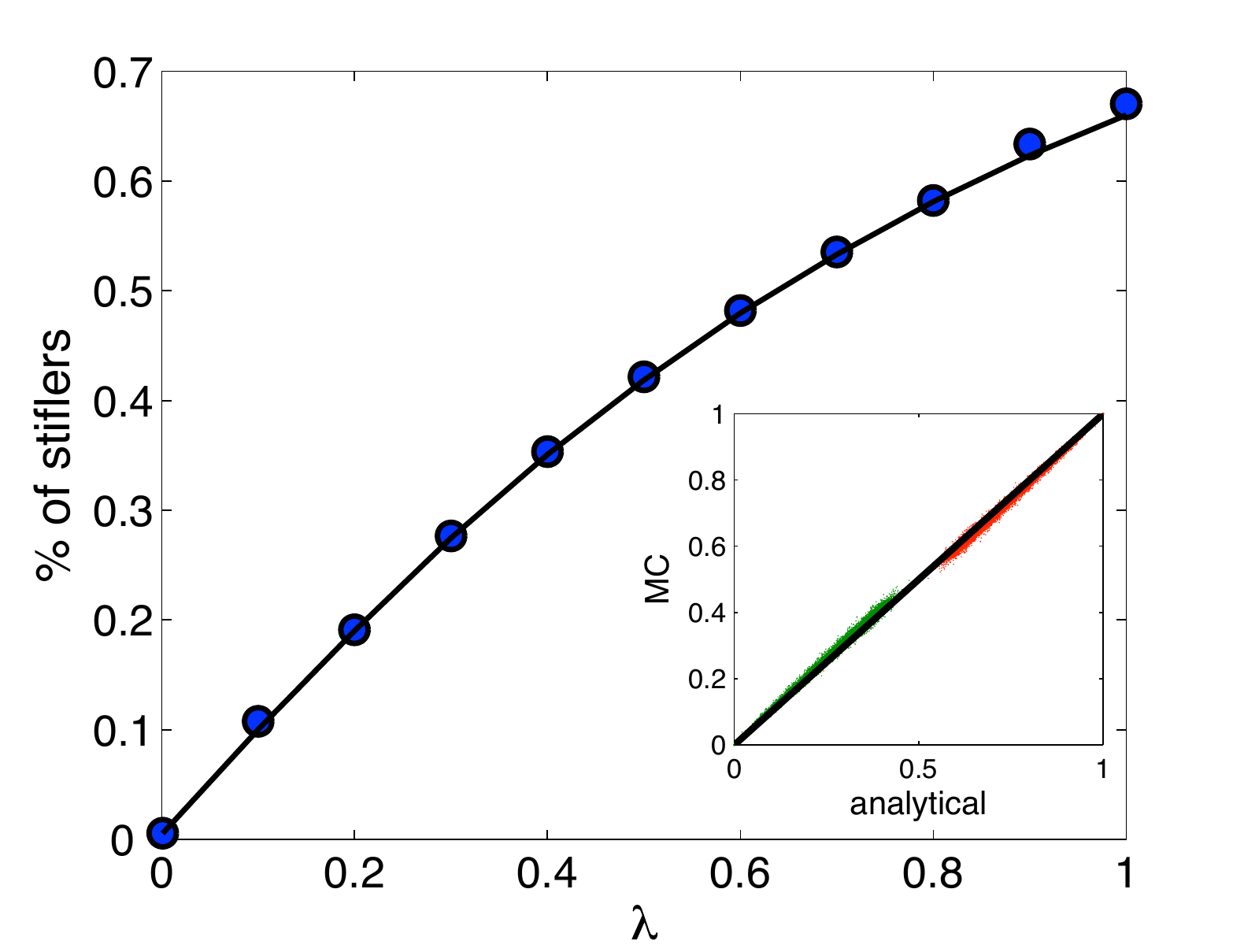}} 
  \caption{(Color online) Homogeneous propagation. Rumor spreading on the top of Barab{\'a}si-Albert (BA) networks with $N = 10^4$ nodes: (a) final density of stiflers (red=1, blue=0) for different spreading rates (dashed white line). Similar values of $\lambda$ yield the same expected density of stiflers;  (b) comparison of the theoretical results (solid line) with Monte Carlo (MC) simulations (symbols). The inset shows the probability that vertex $x$ ($x = 1, \ldots, N$) is a stifler ($\St_x$, red points) or an ignorant ($\Ig_x$, green points) for MC simulations and theoretical predictions. Each simulation is an average over $500$ realizations. }	
\end{figure}

We compare this result with Monte Carlo (MC) simulations. Our simulation considers discrete time steps and begins with a random set of initial seed spreaders. The spreading is performed as a contact process, in which only one spreader is selected at each time step. In the classical homogeneous approach, spreaders infect their ignorant neighbors with the same probability $\beta$ and recover with  probability $\mu$. The spreading rate is defined as $\lambda$ = $\beta / \mu$. In Figure~\ref{fig:lambda} we show the density of stiflers according to different values of $\lambda$ for the homogeneous spreading. We can see that no phase transition is observed in the rumor spreading, unlike the disease propagation~\cite{Pastor-Satorras2015}. Figure~\ref{fig:BA10mil50HOM} compares the theoretical results with Monte Carlo simulations. The simulations start with a uniform selection of $\dSp(0) = 0.01$ initial spreaders. Without loss of generality, we assume $\mu = 1$. The inset  of Figure~\ref{fig:BA10mil50HOM} shows that the theoretical results also capture the microscopic dynamics  of the MC simulations.

\begin{figure*}[!btp]
\centering	
  \subfigure[]{\label{fig:modelDG}\includegraphics[width= 0.32\textwidth]{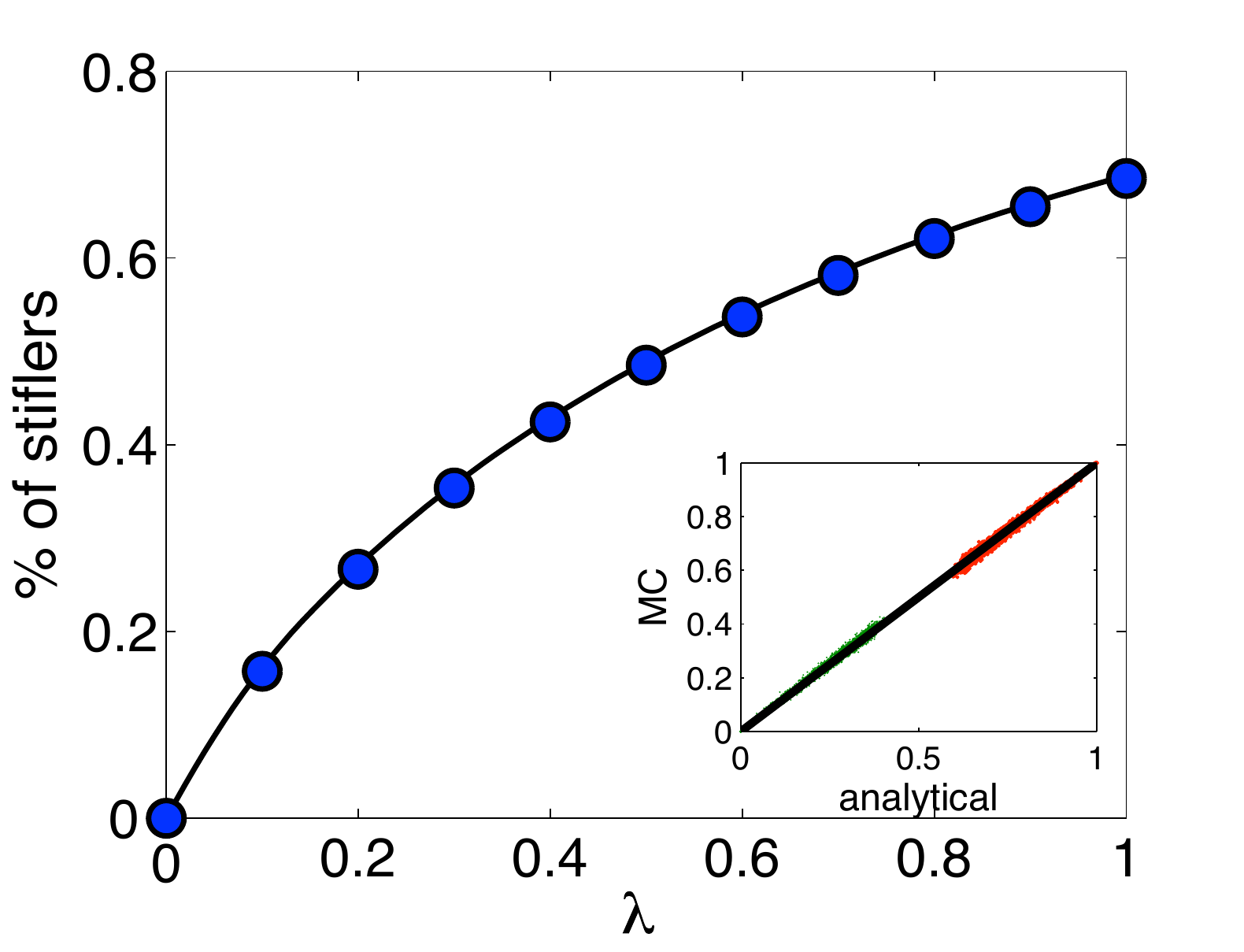}}   
  \subfigure[]{\label{fig:modelBE}\includegraphics[width= 0.32\textwidth]{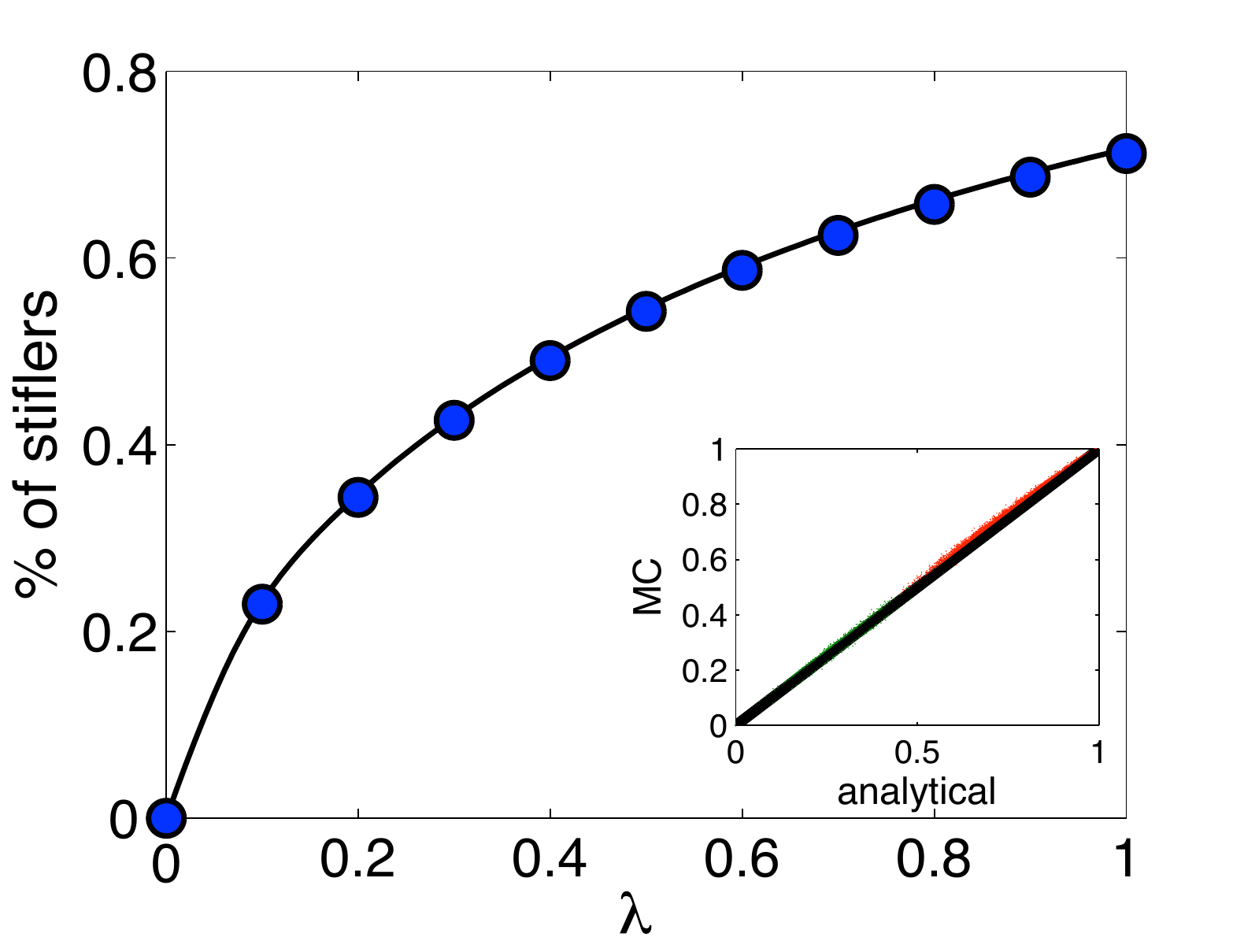}}  
  \subfigure[]{\label{fig:modelPR}\includegraphics[width= 0.32\textwidth]{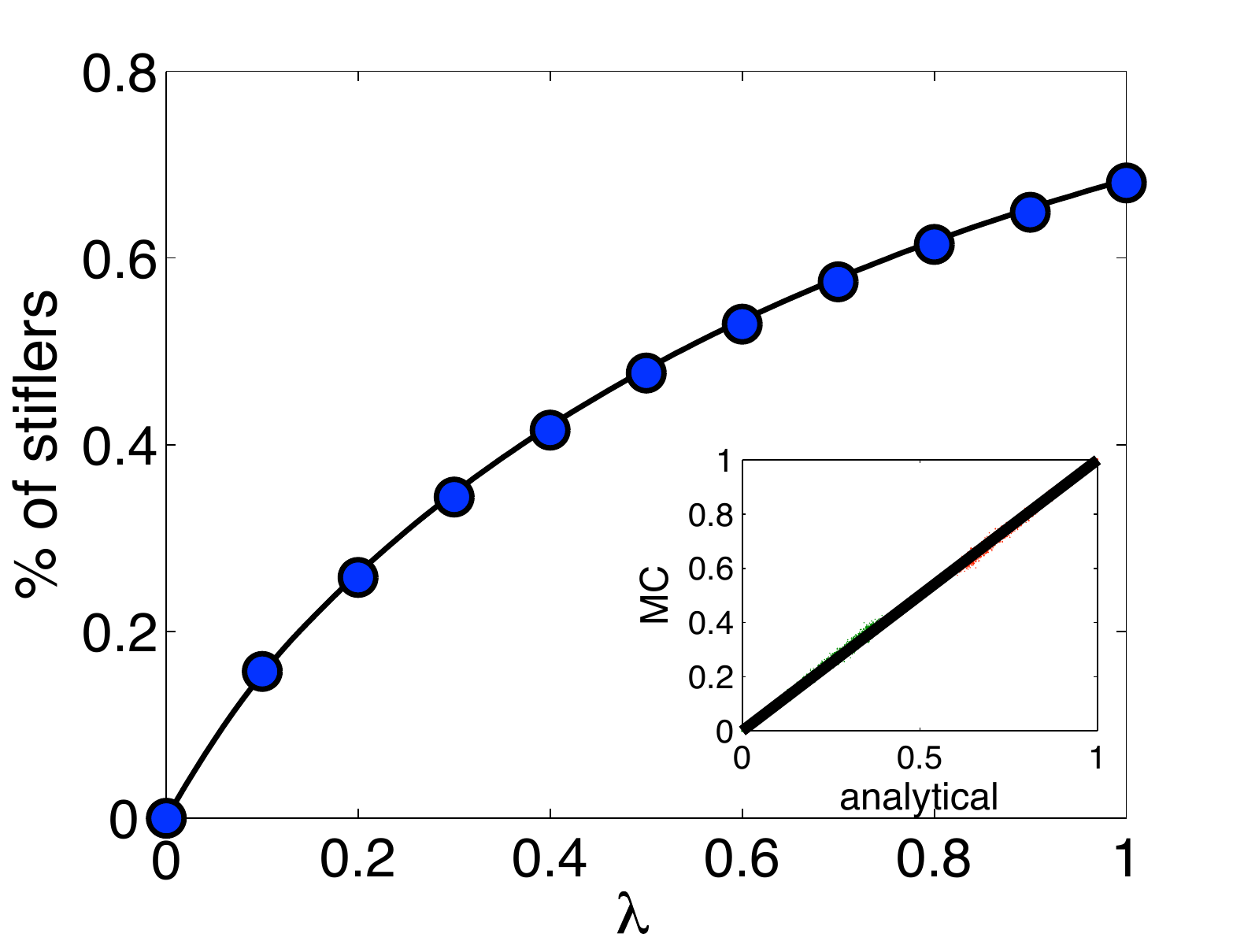}}   
\caption{\label{fig:modelHETDiagram} 
(Color online) Heterogeneous propagation. Final density of stiflers in a Barab{\'a}si-Albert (BA) network of size $N = 10^4$. Symbols correspond to Monte Carlo (MC) simulations whereas solid lines are solutions of the theoretical analysis. The propagation probability $\beta_x$ is completely proportional to the centrality measures: (a) degree, (b) betweenness centrality and (c) PageRank.  The inset shows the probability that vertex $x$ ($x = 1, \ldots, N$) is a stifler ($\St_x$, red points) or an ignorant ($\Ig_x$, green points) for MC simulations and theoretical predictions (Equation~\ref{Eq:Rx}). Each simulation is an average over $500$ realizations.}
\end{figure*}

The results for the heterogeneous propagation in a Barab{\'a}si-Albert (BA) network of size $N = 10^4$ are shown in Figure~\ref{fig:modelHETDiagram}. The propagation probability $\beta_x$ is assumed to be proportional to a centrality measure, i.e., $\beta_x = \alpha_x / \max_{y=1,\ldots, N}(\alpha_y)$, such that $0\leq\beta_x\leq 1$, $x=1,2,\ldots, N$ and $\alpha_x$ is a centrality measure (e.g. degree, betweenness centrality or PageRank). Also, without loss of generality, we assume $\mu_x = 1$. We observe that the analytical solutions agree very well with the results of MC simulation.  As in the homogeneous case, no phase transition is observed. Hence, contrariwise to other dynamical processes, such as synchronization~\cite{Gomez011, Peron012}, the correlation between the dynamical parameters and network properties does not affect the dynamical behavior of the system.

\section{Databases}\label{sec:database}

We explore the heterogeneous propagation in artificial and real-world networks. For artificial networks, we consider the Erd{\"o}s-R{\'e}nyi (ER) random graph model~\cite{erdosrenyi1959}, Watts-Strogatz (WS) small world model ~\cite{watts98}, the Barab{\'a}si-Albert (BA) scale-free model~\cite{barabasiEalbert1999} and the Barth{\'e}lemy spatial model (SSF)~\cite{Barthelemy2003}. Real-world networks comprise collaboration between scientists, email exchanging, and social networks.

\subsection{Network Models} 

The Erd{\"o}s-R{\'e}nyi (ER) model~\cite{erdosrenyi1959} generates random networks assuming that all vertices are statistically similar. A Bernoulli process constructs the network by assigning a connection between each pair of nodes with probability $p$, where self-connections are avoided. The expected number of connections is given by $\left\langle k\right\rangle = (N - 1)p$~\cite{erdosrenyi1959}. In the thermodynamic limit $(N \rightarrow \infty )$ and considering a fixed average degree, the degree distribution follows the Poisson distribution, i.e.,
\begin{equation}
 P(k) = \frac{e^{-\left\langle k\right\rangle} {\left\langle k\right\rangle}^k}{k!} \:.
\label{eq:poisson}
\end{equation} 
The ER network model does not present hubs (densely connected nodes), but a homogeneous organization in which nodes have degree close to the average degree $\left\langle k\right\rangle$~\cite{costa07}. 

The Watts and Strogatz (WS) small world model is a simple model for interpolating between regular and random networks~\cite{watts98}. The network construction begins with a regular ring of $N$ vertices connected with its $k$ nearest neighbors in both direction, $k/2$ neighbors on each side, with $M= kN$ edges. For each vertex, edges are rewired in the clockwise direction with probability $q$. The rewiring is a uniform random selection that avoids self-loops and edges duplication (multi-edges). When $q = 1$, the model generates random networks whose vertices have degree  $k_{x} \geq k$. When $q = 0$, the model produces a ring network with the largest fraction of triangles. 

The Barab{\'a}si-Albert (BA) model~\cite{barabasiEalbert1999} generates scale-free networks with power-law degree distribution, i.e., $ P(k) \sim k^{-\gamma}$, where $\gamma = 3$ in the thermodynamic limit~\cite{barabasiEalbert1999}. The network organization is highly heterogeneous and displays a few hubs and many low-degree nodes. To construct the network, we begin with a initial network made up by $N_0$ vertices. A new vertex  with  $k'$ $(k' < N_0)$  edges is included at each time step $t = \{1,\: 2\:, 3\: ...\: N - N_0\}$ and $N > N_0$. The incoming vertex $x$ connects to node $y$ according to the probability
\begin{equation}
\label{eq:BA}
	\mathcal{P}(x \rightarrow y) \ = \ \frac{k_y}{\sum_{j} \ {k_j}} \:.
\end{equation} 
After $t$ time steps, we obtain a random network with  $t + N_0$ vertices and average degree $\left\langle k \right\rangle \approx 2k'$. Notice that hubs tend to grow more connections than low degree nodes.

The Barth{\'e}lemy (SSF) model~\cite{Barthelemy2003} constructs scale-free networks embed on space. Vertices are uniformly located in a regular $d$ dimensional lattice of size $l$. The model incorporates the concepts of proximity selection, with some distance kernel between vertices, and preferential attachment. The probability of connection of a new incoming vertex $x$ given by 
\begin{equation}
\label{eq:SSF}
	\mathcal{P}(x \rightarrow y) \ \propto \ \frac{k_y +1}{\exp(d_{xy}/r_c)} \:,
\end{equation} 
where $d_{xy}$ is the Euclidean distance between vertices $x$ and $y$. The $r_c$ scale is a finite parameter that controls the degree correlation~\cite{costa07} and the proportion of triangles~\cite{watts98}. The algorithm to generate a network initiates with all vertices uniformly distributed in a two-dimensional space. Next, it is selected $N_0$ initial active vertices at random. An inactive vertex $x$ is taken at random and connected with an active one according to the probability function $\mathcal{P}$. Vertex $x$ becomes active and the two last steps are repeated until all vertices are activated. The average degree of the network is satisfied by repeating the second step $k'$ times for each vertex, which results in $\left\langle  k\right\rangle = 2k'$. This model can generate scale-free or homogeneous networks embedded in space according to the values of the parameters $r_c, L$ and $d$. Here we consider values $r_c = 0.05$, $L = 1$ and $d=2$, producing scale-free networks, as reported in the original paper~\cite{Barthelemy2003}.

\subsection{Social networks}

\emph{Netscience} is a coauthorship network of scientists working on Network Science. The data were compiled from the references of the article reviews by Newman~\cite{newman2003} and Boccaletti et al.~\cite{boccaletti06}. Nodes represent authors and two authors are connected by an edge if they have coauthored one or more papers. The data form an unweighted and undirected network with $1{,}590$ vertices and $2{,}742$ edges. 

The \emph{email} network represents the e-mail interchanges between members of the Univeristy Rovira i Virgili, Tarragona~\cite{Guimera03}. Although it is originally a directed network, we assume it as an undirected one with $N= 1{,}133$ vertices and $M= 5{,}451$ edges. 

\emph{Hamsterster} is an undirected and unweighted network based on data from \textsc{hamsterster.com} website~\cite{konectURL}. The vertices represent users of the system, while the edges correspond to friend and family relationship. The network has $2{,}426$ vertices and an average degree of $13.71$. 

\emph{Google+} is a directed and unweighted social network that contains user-user links~\cite{GplusPaper}. The directed edge denotes if one user added another one in his circles. Here we assume the network as undirected. The network has $23{,}613$ vertices and $M= 39{,}183$ edges, yielding an average degree of $3.31$. 

Structural measures of these social networks are presented in Table~\ref{tab:measures}. Only the main component of the networks is considered here. We see that most networks are sparse and have a very short average path length.

\begin{table*}[!ht]
	\centering
	\caption{Structural properties of the complex networks: number of vertices ($N$), mean degree ($\left\langle k \right\rangle$), average shortest path length ($\left\langle \ell \right\rangle$), average betweenness centrality ($\left\langle b_x \right\rangle$), average closeness centrality ($\left\langle c_x \right\rangle$), average eigenvector centrality ($\left\langle X_x \right\rangle$), $K$-core ($\left\langle Kc(x) \right\rangle$) and PageRank ($\left\langle \pi_x \right\rangle$).}
		\begin{tabular}{ c c c c c c c c c c c }
		\hline
 	&Network & $N$ & $\left\langle k \right\rangle$ & $\left\langle \ell \right\rangle$  & $\left\langle b_x \right\rangle$ & $\left\langle c_x \right\rangle$ & $\left\langle X_x \right\rangle$ & $\left\langle Kc(x) \right\rangle$ & $\left\langle \pi_x \right\rangle$\\ \hline

	& ER & $1{,}000$ & $12$ & $3.04$ &  $1.02\times 10^3$ & $3.30\times 10^{-4}$ & $0.030$ & $7.81$ & $10^{-3}$ \\	
Artificial		& WS & $1{,}000$ & $12$ & $3.59$ & $1.69\times 10^3$ & $2.29\times 10^{-4}$ & $0.031$ & $5.98$ & $10^{-3}$ \\
	& BA & $1{,}000$ & $12$ & $2.84$  & $9.20\times 10^2$ & $3.54\times 10^{-4}$ & $0.022$ & $6.00$ & $10^{-3}$ \\		
	& SSF& $1{,}000$ & $12$ & $4.08$ & $1.54\times 10^3$ & $2.48\times 10^{-4}$ & $0.022$ & $6.08$ & $10^{-3}$ \\	
	\hline		
	&\emph{netscience} & $379$  &  $4.82$ & $6.04$ & $9.53\times 10^2$ & $4.51\times 10^{-4}$ & $0.016$ & $3.47$ & $2.64\times 10^{-3}$ \\
Social 	&\emph{email} & $1{,}133$ &  $9.62$ & $3.60$ & $1.48\times 10^3$ & $2.49\times 10^{-4}$ & $0.017$ & $5.35$ & $8.83\times 10^{-4}$ \\	
	&\emph{Hamsterster} & $2{,}000$ & $16.1$ & $3.59$ & $2.59\times 10^3$ & $1.43\times 10^{-4}$ & $0.011$ & $9.29$ & $5.00\times 10^{-4}$ \\
	&\emph{Google+} & $23{,}613$ & $3.32$ & $4.03$ & $3.58\times 10^4$ & $0.11\times 10^{-4}$ & $0.002$ & $1.67$ & $4.21\times 10^{-5}$ \\
	\hline
		\end{tabular}			
	\label{tab:measures}
\end{table*}

\section{Simulations and Results}\label{sec:simulation}

We perform Monte Carlo simulations to analyze the influence of heterogeneous probability distribution on the spreading dynamics. The contact process is adopted in our simulations, i.e., only one spreader is selected to propagate the information at each time step. We define the spreading capacity of node $x$ as the final density of informed individuals (stiflers) obtained when the simulation starts on $x$, i.e., when $x$ is selected as the initial seed spreader. The results are obtained by averaging over $30$ simulations.

In order to compare the impact of heterogeneous configuration against the classical case, we assume 
\begin{equation}
 \beta_x = p \beta + (1  - p) \theta_x \: , \hbox{ with } 0 \leq p \leq 1,
\label{eq:hetri2}
\end{equation} 
to define the transmission probabilities of each node $x$. In this way, we can control the level of heterogeneity in information propagation by setting the parameter $p$ between zero and one. The heterogeneous propagation occurs when $ p < 1$, whereas $p = 1$ recovers the classical homogeneous case, in which all nodes have the same capacity of information spreading. Moreover, since centrality measures are defined in different scales, we consider their normalized version, i.e., $\theta_x = \alpha_x / \max_{y=1,\ldots, N}(\alpha_y)$, such that $0\leq\theta_x\leq 1$, $x=1,2,\ldots, N$ and $\alpha_x$ is a centrality measure. Centrality measures are presented as an appendix at the end of this paper. Finally, we keep the average probability of propagation $\beta$ as constant independent of $p$, i.e.,  $\beta=\sum_{x=1}^N \beta_x/N$. This assumption is necessary to perform an appropriate comparison between the homogeneous and heterogeneous cases.

\subsection{Network models}

\begin{figure*}[!ht]
\centering
\subfigure[]{\label{fig:ER1000TC}\includegraphics[width= 0.45\textwidth]{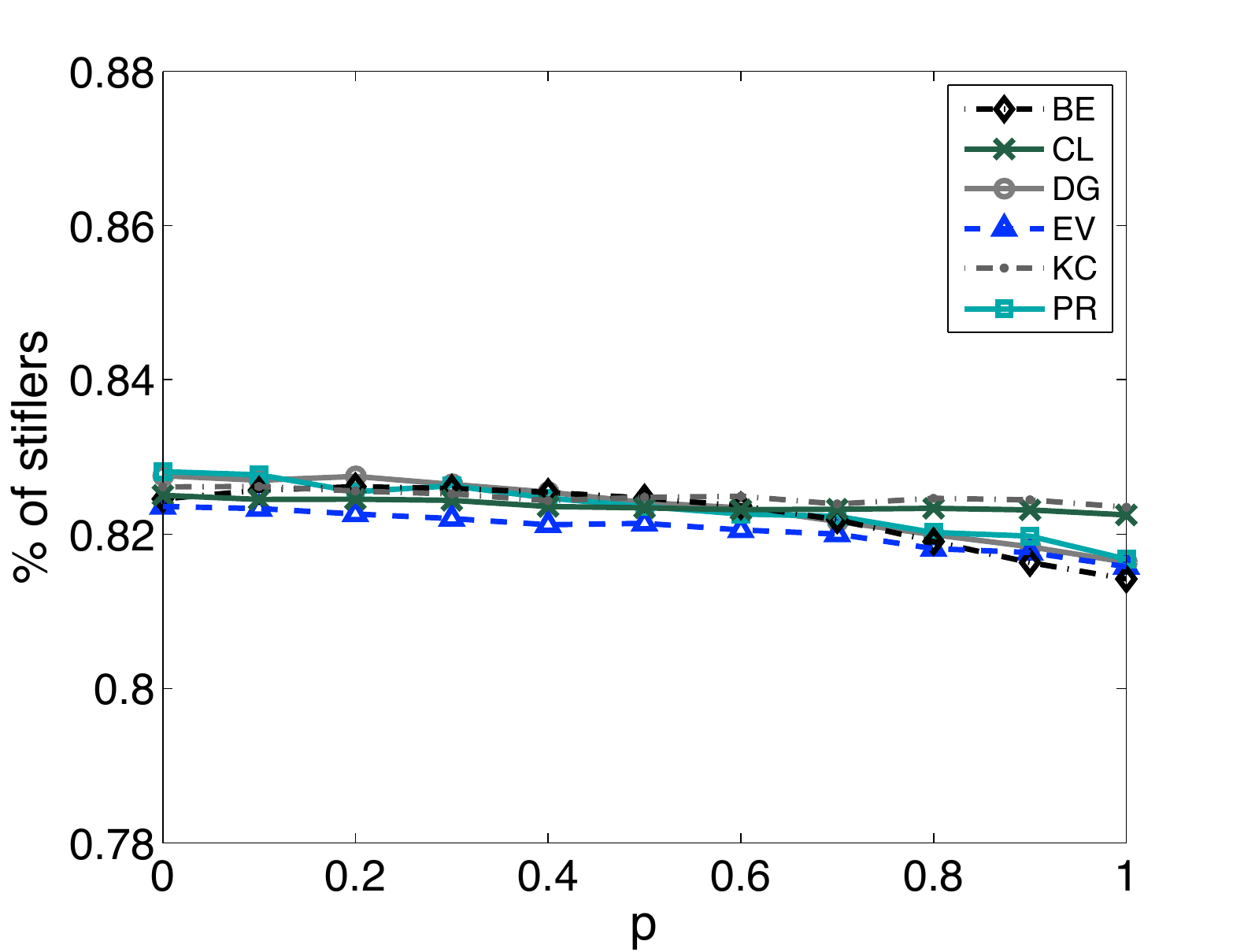}}
\subfigure[]{\label{fig:WSCPaper}\includegraphics[width= 0.45\textwidth]{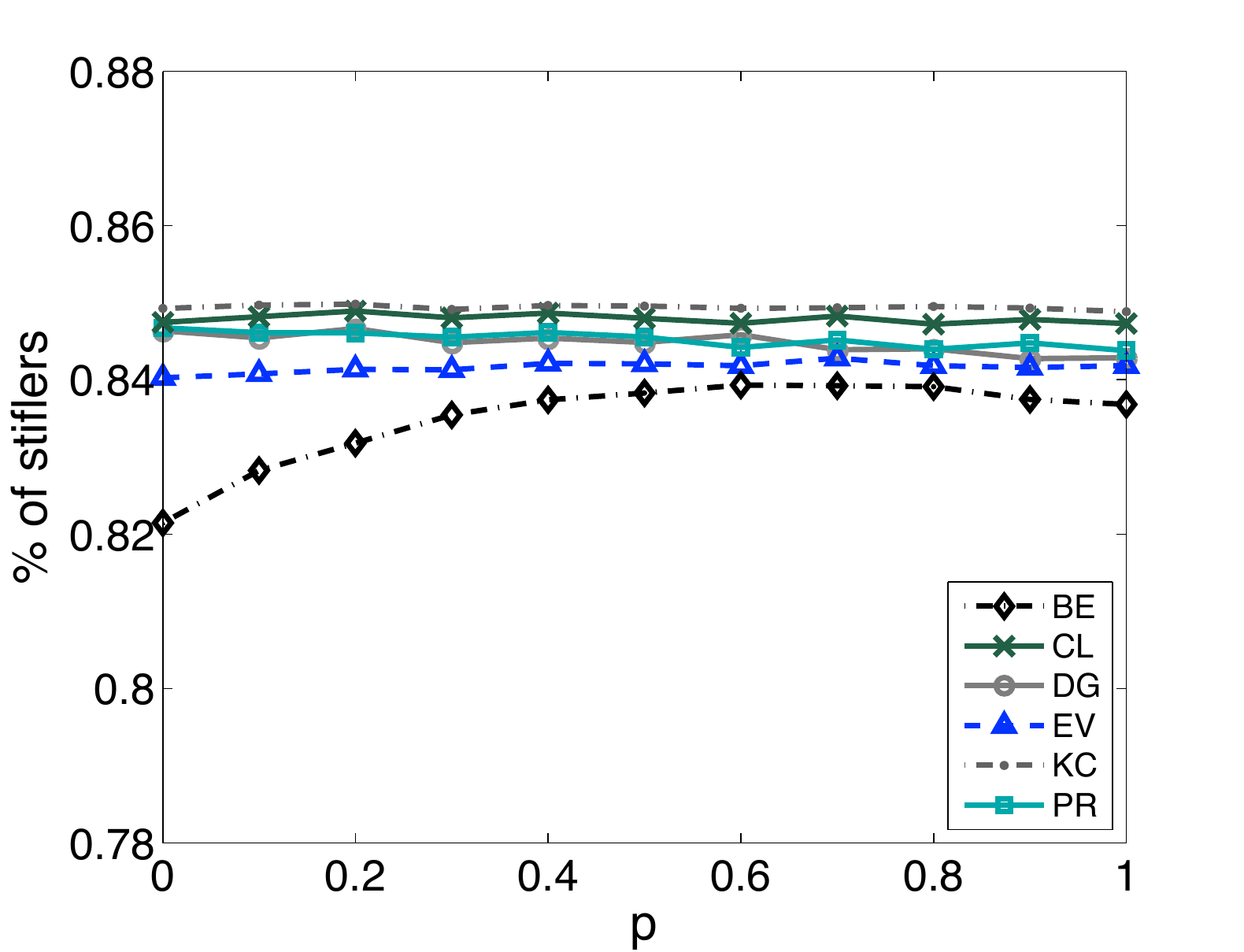}}
\subfigure[]{\label{fig:BA1000CPaper}\includegraphics[width= 0.45\textwidth]{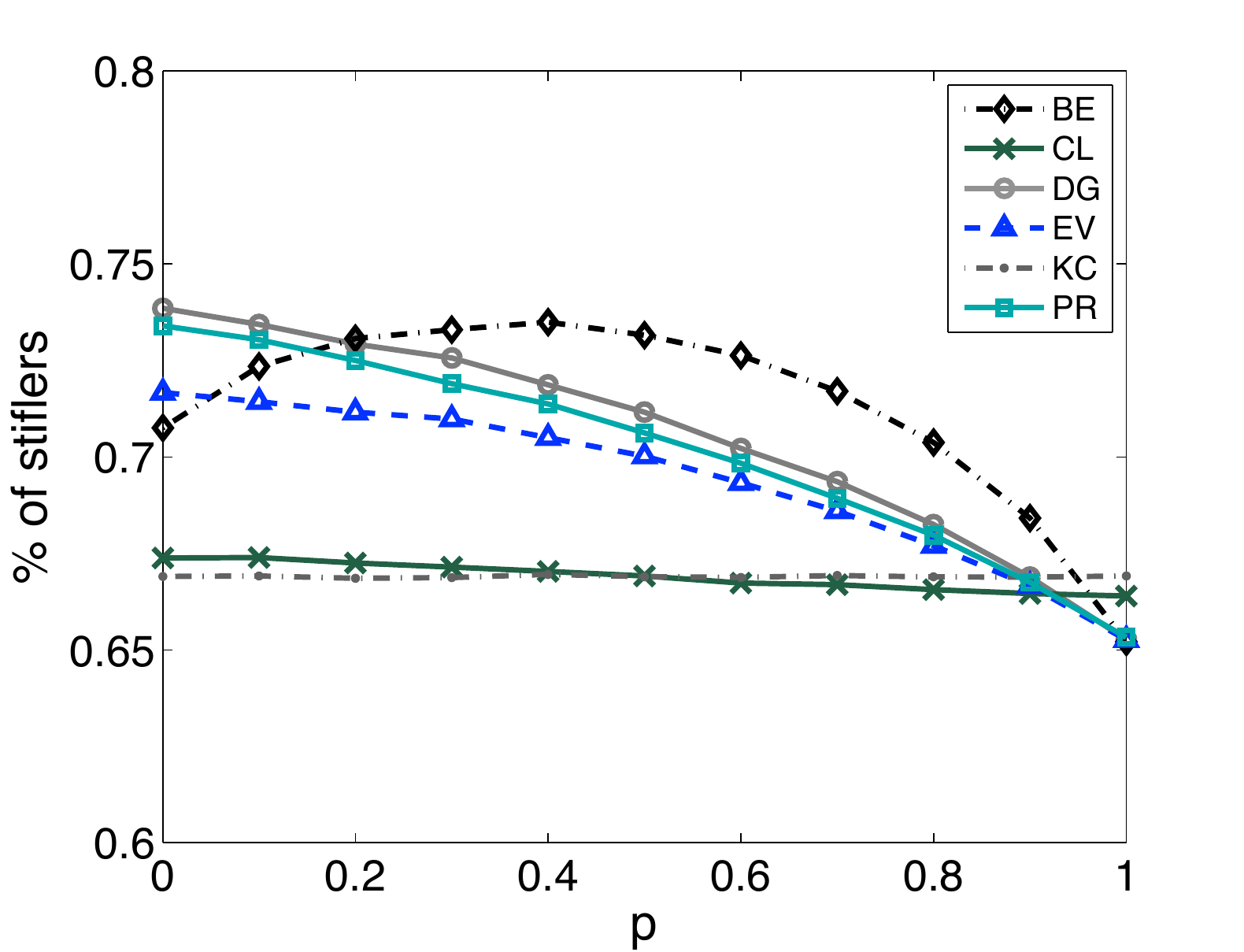}}
\subfigure[]{\label{fig:SSFCPaper}\includegraphics[width= 0.45\textwidth]{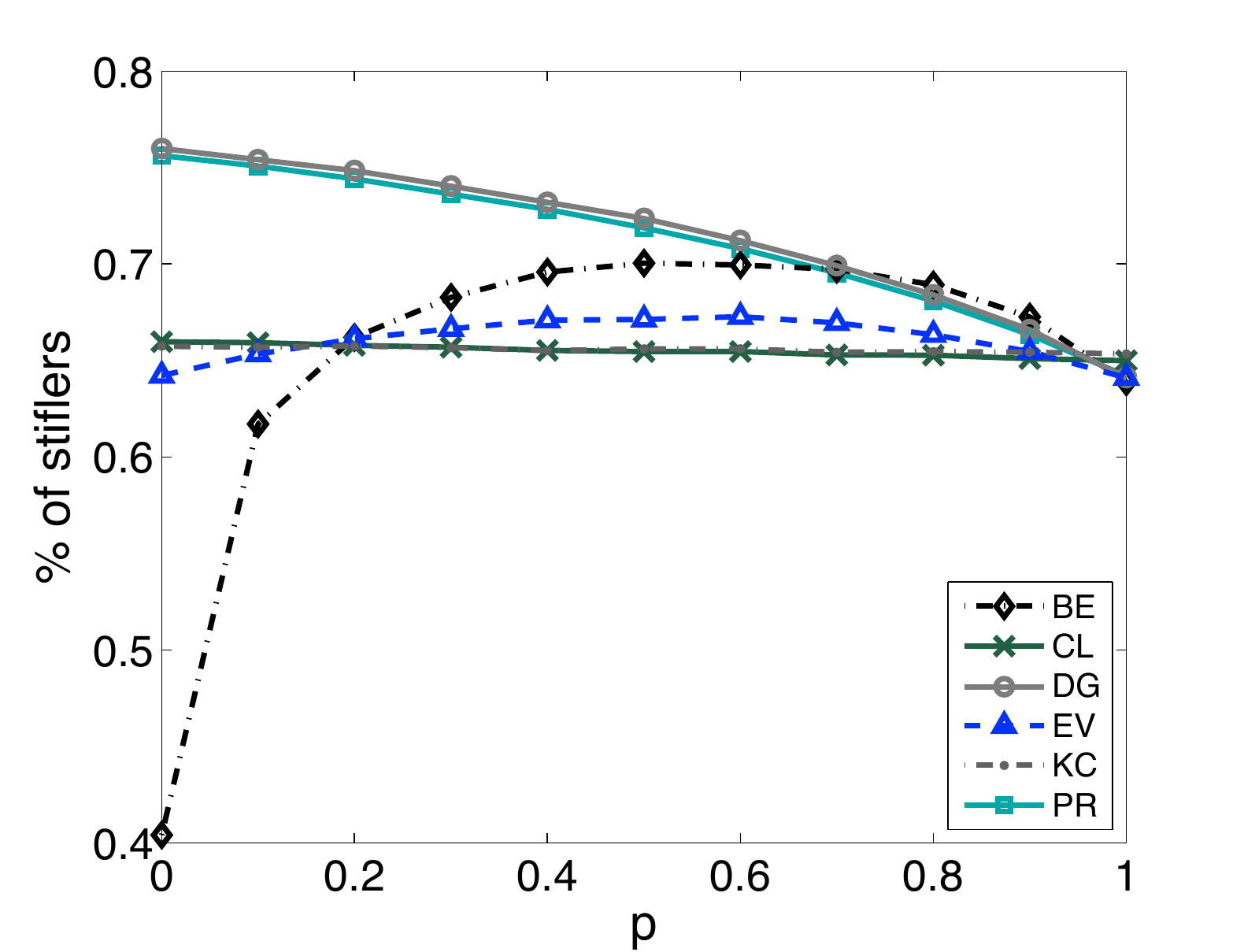}}
\caption{\label{fig:ArtificalCPaper}
(Color online) Final density of stiflers according to the level of heterogeneity defined by the interpolation function (Eq.~\ref{eq:hetri2}). The centrality measures considered are betweenness centrality (BE), closeness centrality (CL), degree (DG), eigenvector centrality (EV), k-core (KC) and PageRank (PR). The homogeneous propagation is given for $p=1$, whereas the completely heterogeneous case occurs for $p=0$. The simulations are performed on the top of  (a) Erd{\"o}s-R{\'e}nyi (ER), (b) Watts-Strogatz (WS), (c) Barab{\'a}si-Albert (BA) and (d) spatial scale-free (SSF) networks.  The spreading rate $\lambda = \sqrt{2}$ is adopted in all simulations.} 
\end{figure*}

\begin{figure*}[!ht]
\centering
\subfigure[]{\label{fig:ERCPaperP}\includegraphics[width= 0.45\textwidth]{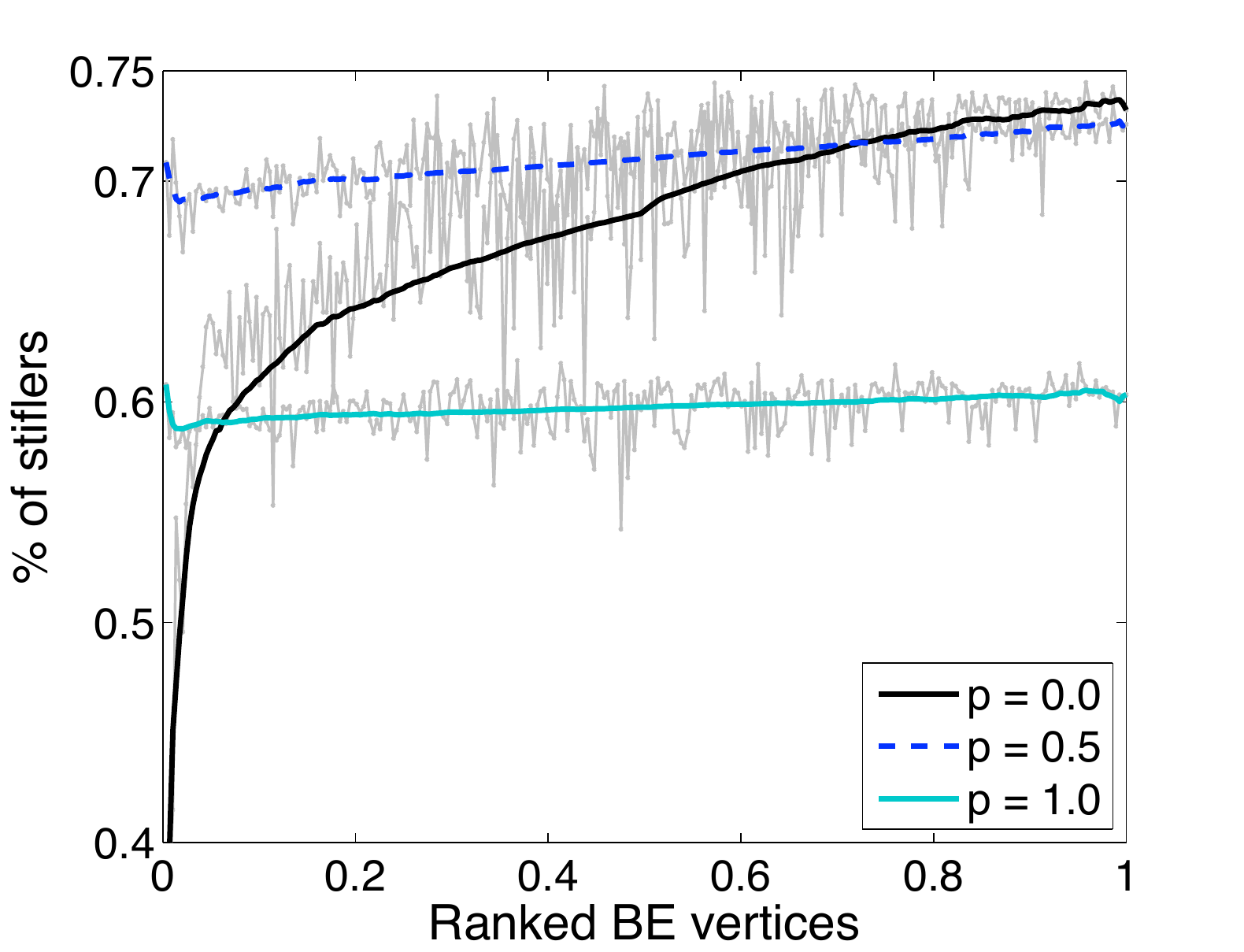}}
\subfigure[]{\label{fig:WSCPaperP}\includegraphics[width= 0.45\textwidth]{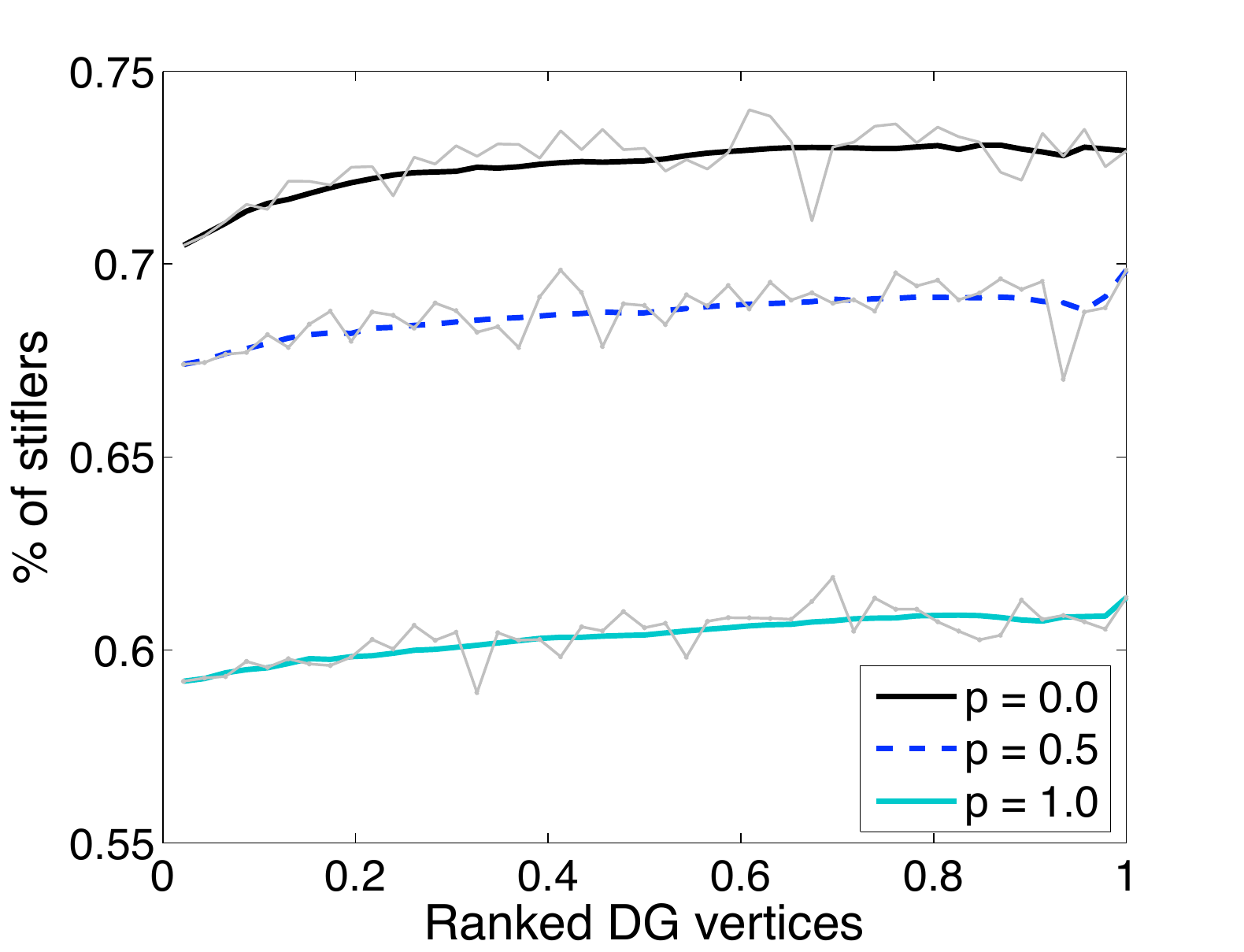}}
\caption{\label{fig:Bet-Dg} 
(Color online) Final fraction of stiflers obtained for each vertex in the Barab{\'a}si-Albert (BA) network, ranked in the x-axis according to (a) betweenness centrality and (b) degree. For $p=0$, we have the heterogeneous propagation, whereas the homogeneous case is obtained for $p=1$. The partial heterogeneous configuration ($p=0.5$) provides the best configuration for rumor propagation in the case of the betweenness centrality. The curves represent the interpolation of the points by least squares fitting technique.}
\end{figure*}

To verify how the network structure influence the rumor propagation, we use theoretical networks generated by the random graph of Erd{\"o}s-R{\'e}nyi (ER), small-world network model of Watts-Strogatz (WS), scale-free model of Barab{\'a}si-Albert (BA) and the spatial scale-free (SSF) network model of Barth{\'e}lemy (SSF). These theoretical networks are generated with the same number of nodes and edges since these quantities are related to the spreading process --- for instance, the spreading is faster on denser networks~\cite{Pastor-Satorras2015}. We assume a fix recovery probability  
$\mu = \beta/ \sqrt{2}$
to perform the simulations, without loss of generality --- in Figure \ref{fig:ArtificalCPaper} we present the final density of spreaders for other values of $\mu$, where $\lambda=\mu/\beta$ and $\beta = \sum_x \beta_x/N$ and $\beta_x$ is given by Eq.~\ref{eq:hetri2}. We see in Figure~\ref{fig:ArtificalCPaper} that the heterogeneous propagation does not impact significantly the final fraction of stiflers in random graphs and small-world networks. This result was expected since these networks present nodes with a similar level of centrality due to the homogeneity of their organization. 
However, in scale-free networks, the heterogeneous propagation affect the final fraction of stiflers when the propagation probability $\beta_x$ is proportional to the degree, betweenness centrality, PageRank and eigenvector centrality. Particularly, for the degree, PageRank and eigenvector centrality, the completely heterogeneous propagation provides the best rumor spreading. 
Regarding the closeness centrality and the $k$-core, the final fraction of stiflers is not very affected by the heterogeneous propagation. This occurs because these measures yield close spacing values~\cite{newman2010networks}. Thus, nodes present similar transmission probability even in the completely heterogeneous case. Comparing all the measures, the probability of propagation proportional to the degree and PageRank yields the highest fraction of informed individuals in the long run.

\begin{figure*}[!t]
\centering	
  \subfigure[]{\label{fig:shuER}\includegraphics[width= 0.446\textwidth]{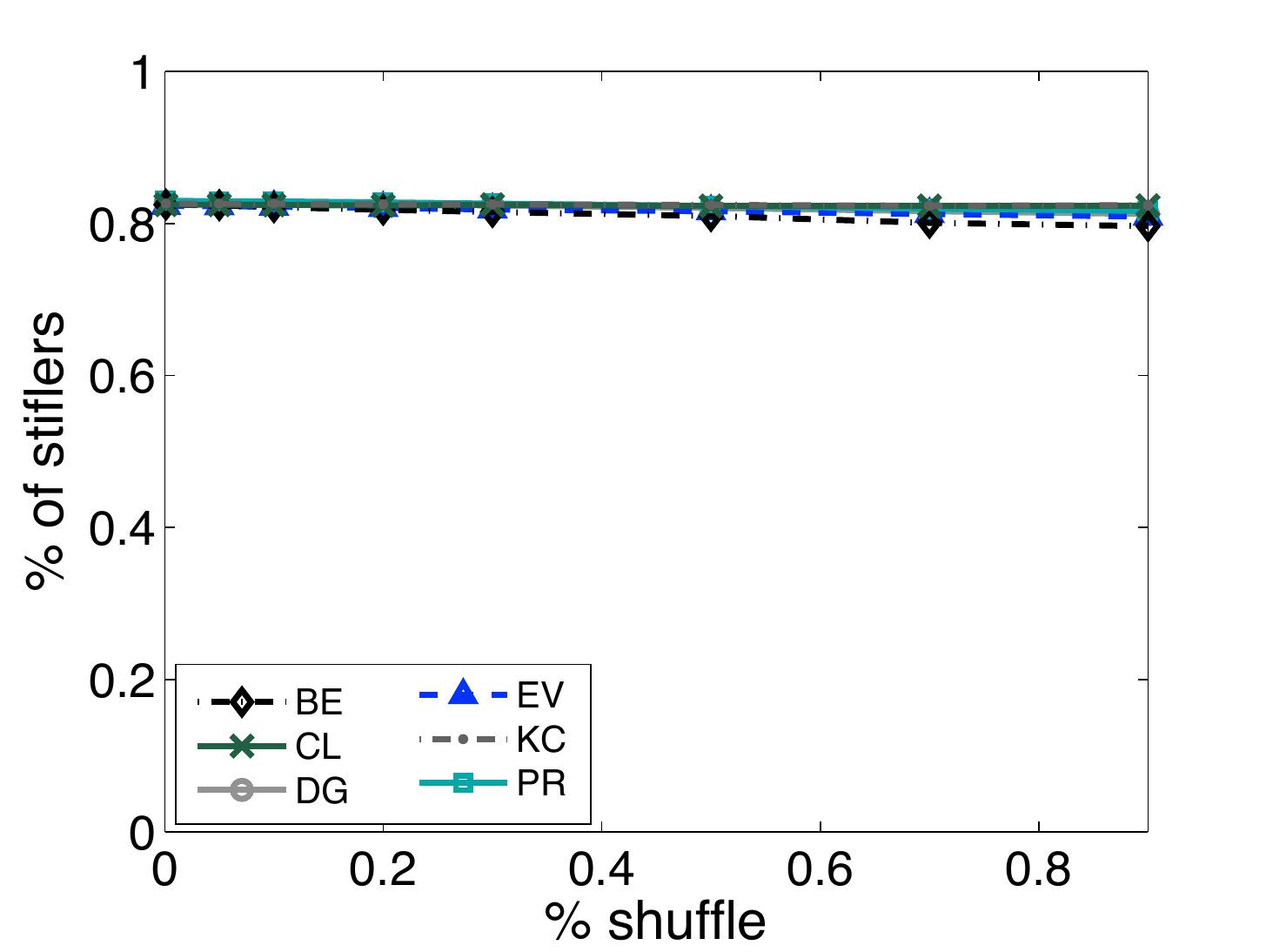}}    
  \subfigure[]{\label{fig:shuWS}\includegraphics[width= 0.446\textwidth]{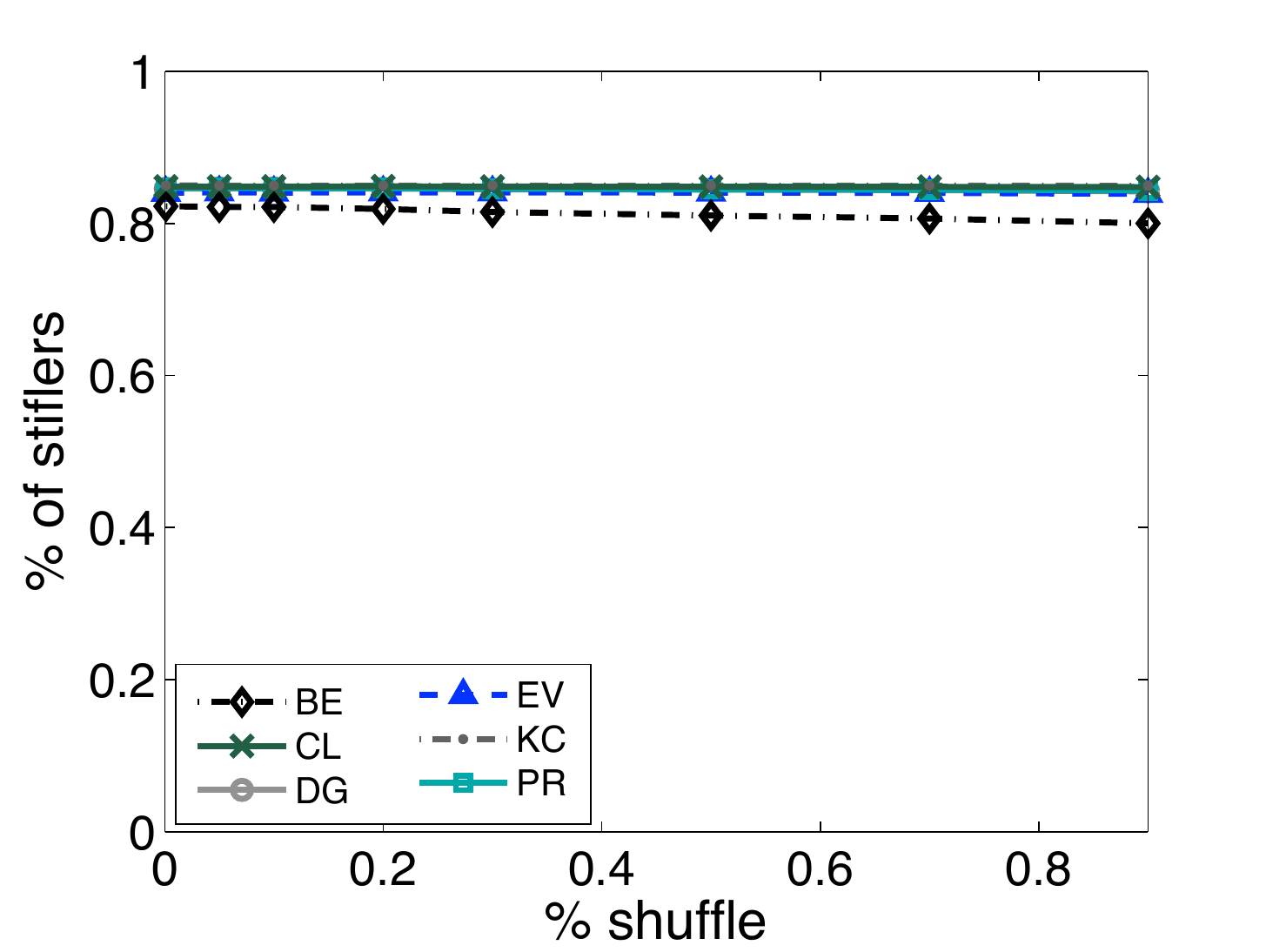}} 
  \subfigure[]{\label{fig:shuBA}\includegraphics[width= 0.445\textwidth]{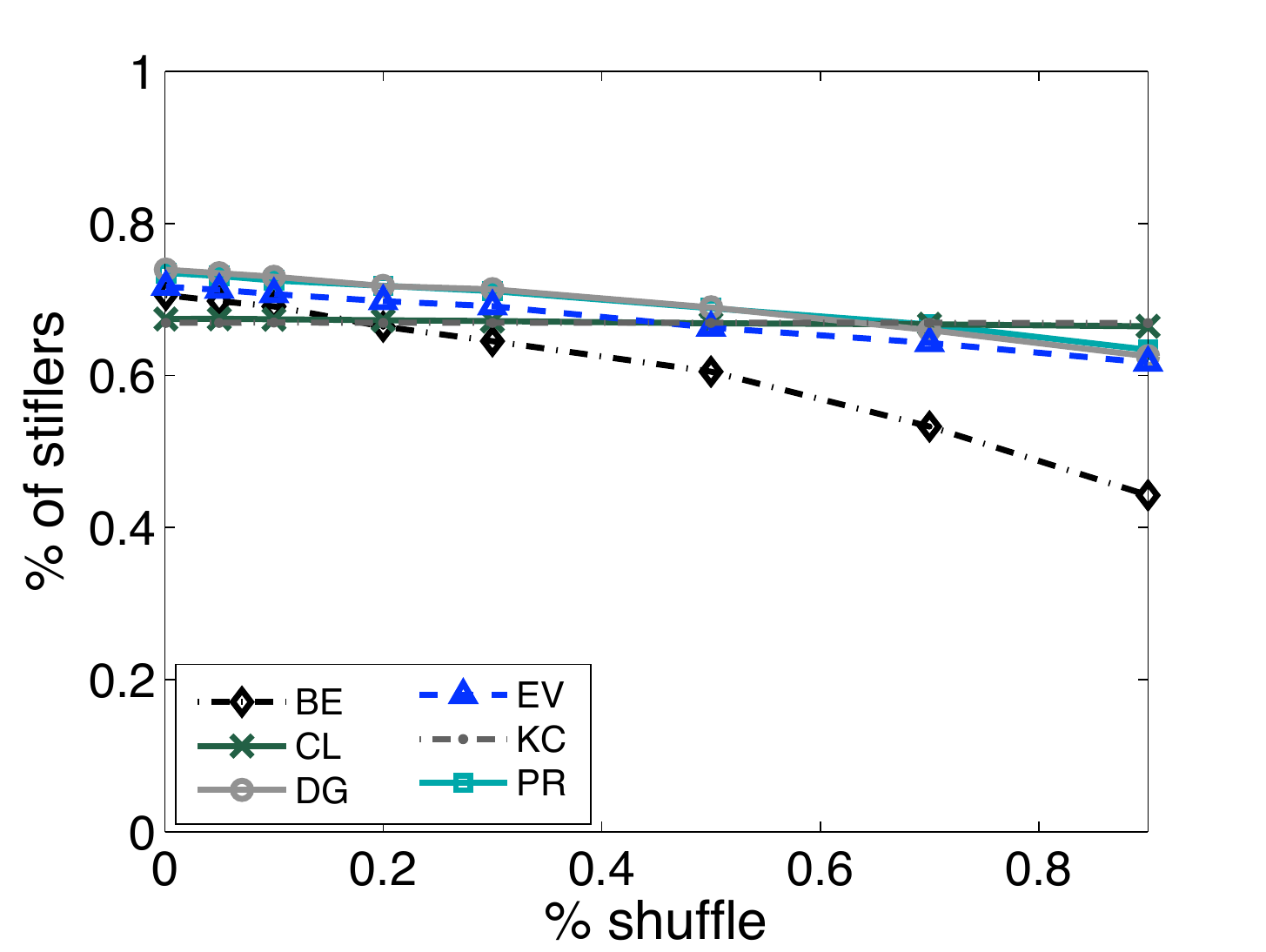}}    
  \subfigure[]{\label{fig:shuSFG}\includegraphics[width= 0.445\textwidth]{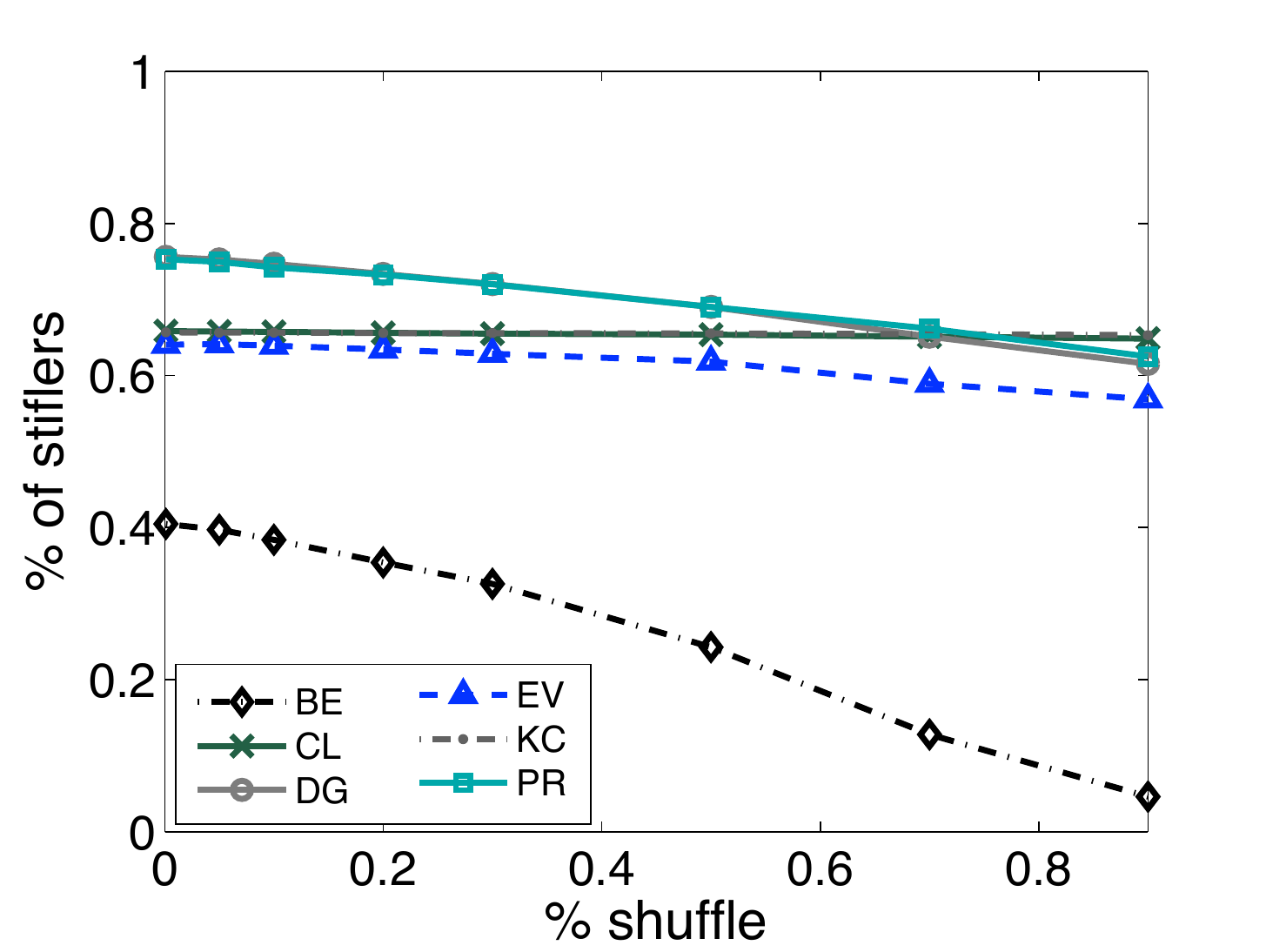}}  
\caption{\label{fig:shuffle} \small (Color online) Final density of stiflers as a function of the percentage of shuffling in the networks generated by the models:  (a) Erd{\"o}s-R{\'e}nyi (ER), (b) Watts-Strogatz (WS), (c) Barab{\'a}si-Albert (BA) and (d) spatial scale-free (SSF). The centrality measures are betweenness centrality (BE), closeness centrality (CL), degree (DG), eigenvector centrality (EV), k-core (KC) and PageRank (PR).  }
\end{figure*}

In the case of the betweenness centrality, the maximum number of stiflers is reached for $p \approx 0.5$. This behavior is explained in Figure~\ref{fig:Bet-Dg}(a). For $p=0$, i.e, when the probability of transmission is proportional to betweenness centrality, the percentage of stiflers obtained from the central nodes is much higher than from peripheral vertices. As the level of heterogeneity is decreased, the spreading capacity of central vertices are reduced, but it is increased the probability of spreading to peripheral nodes. For $p\approx 0.5$, we have the best configuration to obtain the highest number of informed users, since the spreading capacity of central nodes is somewhat decreased, but it is highly increased the capacity of the peripheral nodes. For higher values of $p$, the spreading probability of the most central nodes is too reduced, although the probability of propagation of the peripheral nodes are enhanced, and the average reach of the rumor is lowered. This behavior is not observed for other centrality measures. As shown in Figure~\ref{fig:Bet-Dg}(b) for the degree, the complete heterogeneous propagation provides the best configuration for information spreading. 

Since the correlation between node centrality measures and their propagation probability improve the rumor spreading, it is important to verify the impact of such correlation. We consider the completely heterogeneous case ($p = 0$), i.e., the probability of each node is  proportional to a centrality measure. Then, we perform shuffling in these probabilities to decrease the correlation between the structural and dynamic parameters. Before start each simulation, we select a percentage of pairs of nodes at random and change their spreading probabilities. That is, if we select nodes $x$ and $y$, then we make the spreading probability of $x$ equal to $\beta_y$ and the probability of $y$ as $\beta_x$. After that, the simulation is conducted with the shuffled propagation probabilities. In Figure~\ref{fig:shuffle}, we  observe that for ER and WS networks, the fraction of shuffling has no impact on the rumor spreading. On the other hand, the shuffling affects the final fraction of stiflers in scale-free networks. For the betweenness centrality, degree, PageRank, and eigenvector centrality, the fraction of informed individuals decreases when the shuffling rate is increased. Thus, the correlation between node centrality properties and its probability of information propagation is important to improve the spreading in scale-free networks.

\subsection{Social networks}

\begin{figure}[!t]
\centering	
  \subfigure[]{\label{fig:BA_DG}\includegraphics[width= 0.45\textwidth]{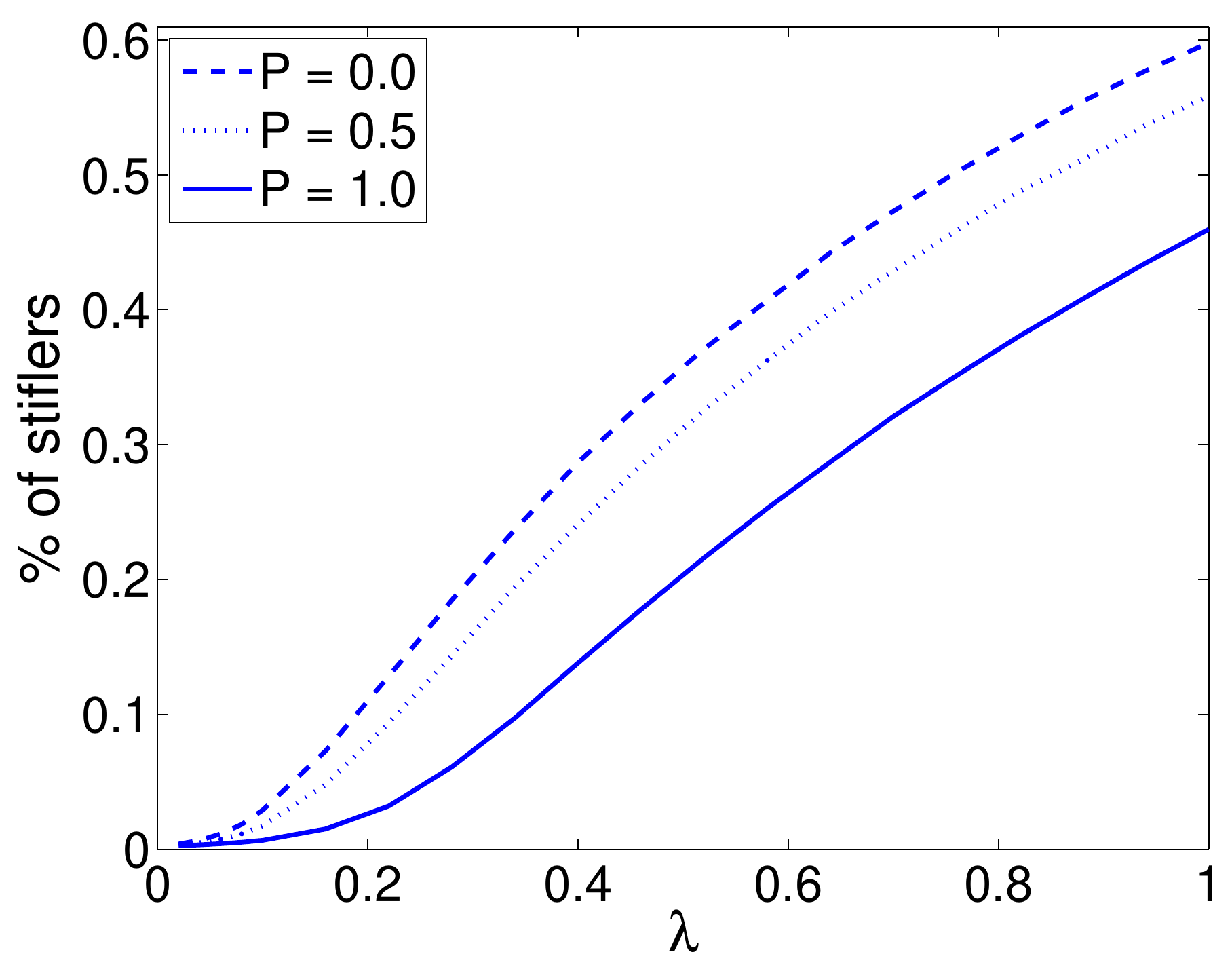}}    
  \subfigure[]{\label{fig:hamsterBC_DG}\includegraphics[width= 0.45\textwidth]{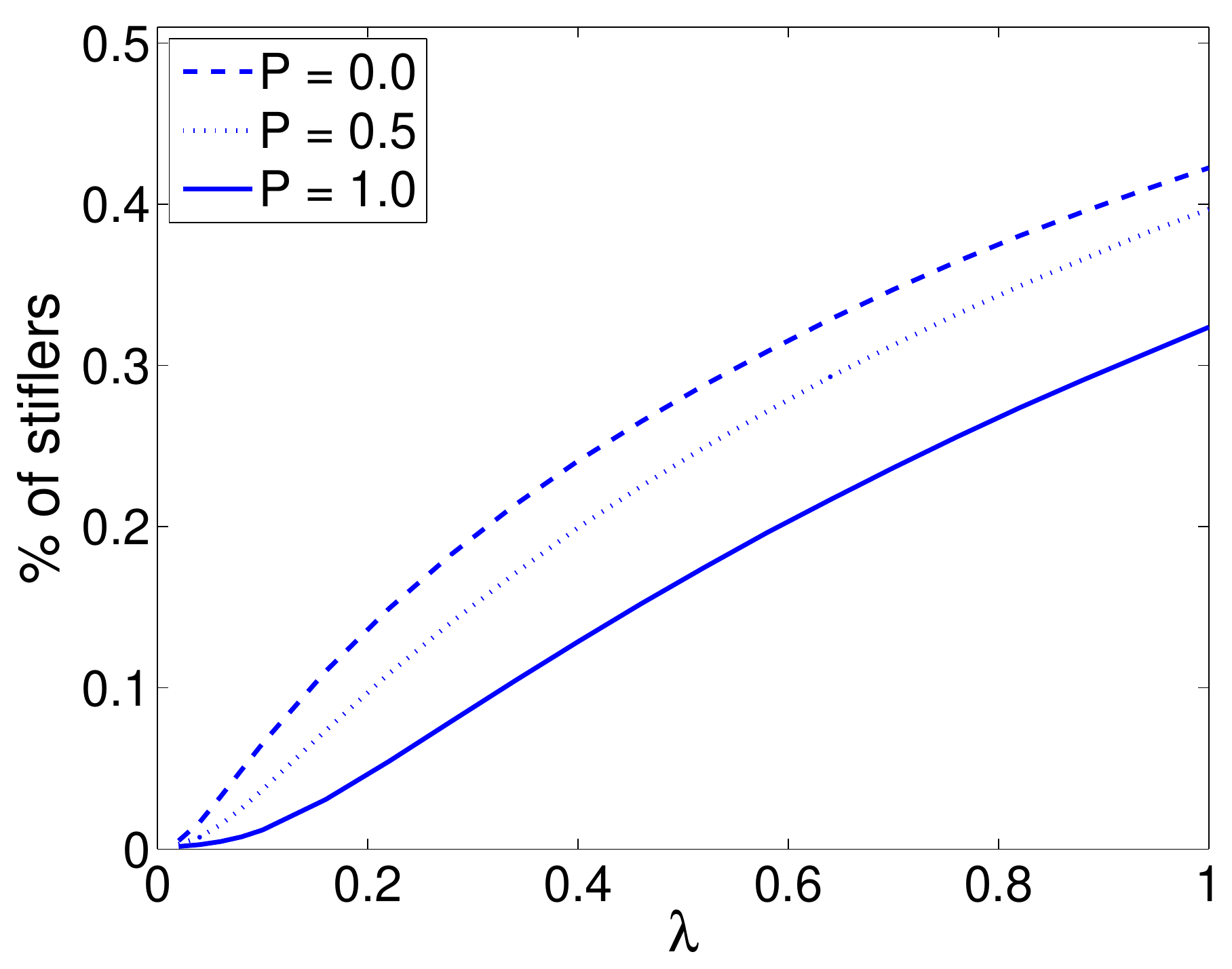}} 
\caption{\label{fig:phase} 
(Color online) Phase diagram of the final fraction of stiflers as a function of parameter $\lambda$ for (a) Barabási-Albert (BA) scale-free networks and (b) \emph{Hamster} network. The heterogeneous propagation considers the degree centrality measure. The homogeneous propagation is given for $p=1$, whereas the completely heterogeneous case occurs for $p=0$.
}
\end{figure}

The simulation of the rumor model is also performed in social networks. Initially, we verify that the heterogeneous propagation does not change the rumor dynamics on social networks since no phase transition is observed, as shown in Figure~\ref{fig:phase}. This behavior is qualitatively similar to that verified in scale-free networks (see the last section and Figure~\ref{fig:BA10mil50HOM}).  When the propagation probability is completely correlated with the degree centrality, the heterogeneous propagation provides the highest reach of information diffusion, as shown in Figure~\ref{fig:phase}. 

\begin{figure*}[!th]
\centering
	\subfigure[]{\label{fig:netscience}\includegraphics[width= 0.42\textwidth]{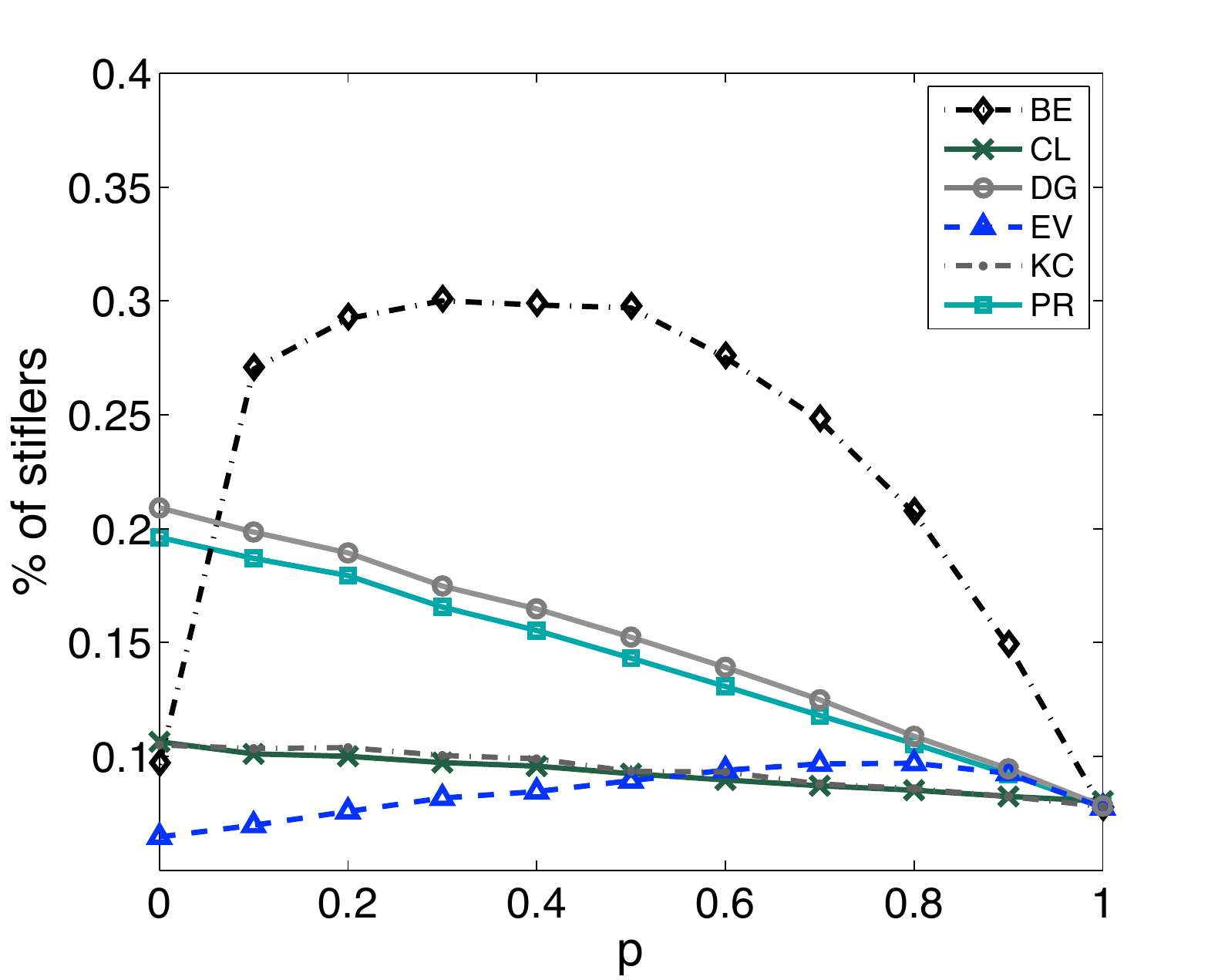}}  
	\subfigure[]{\label{fig:email2}\includegraphics[width= 0.42\textwidth]{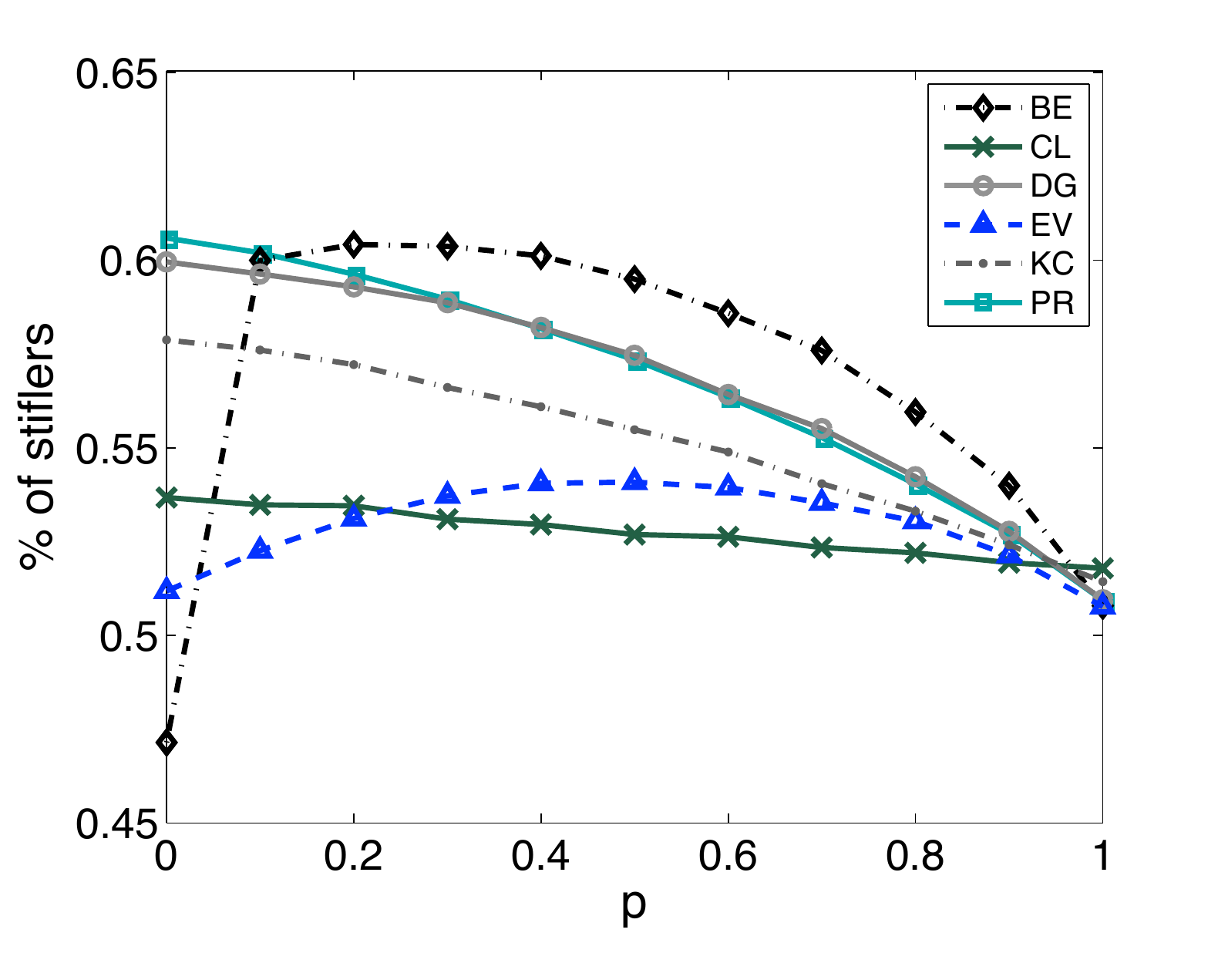}}  
  \subfigure[]{\label{fig:hamster}\includegraphics[width= 0.42\textwidth]{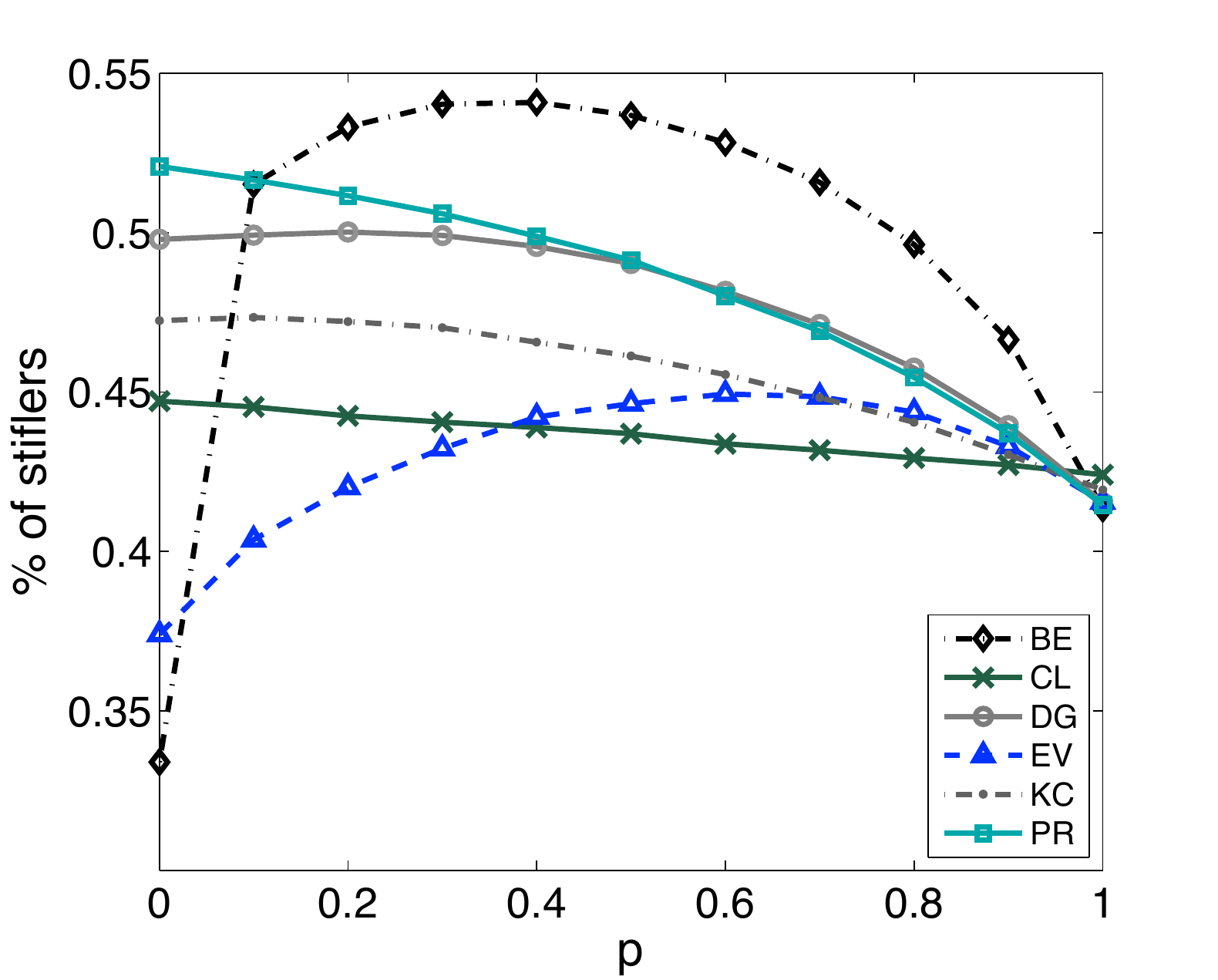}}  
  \subfigure[]{\label{fig:Gplus}\includegraphics[width= 0.42\textwidth]{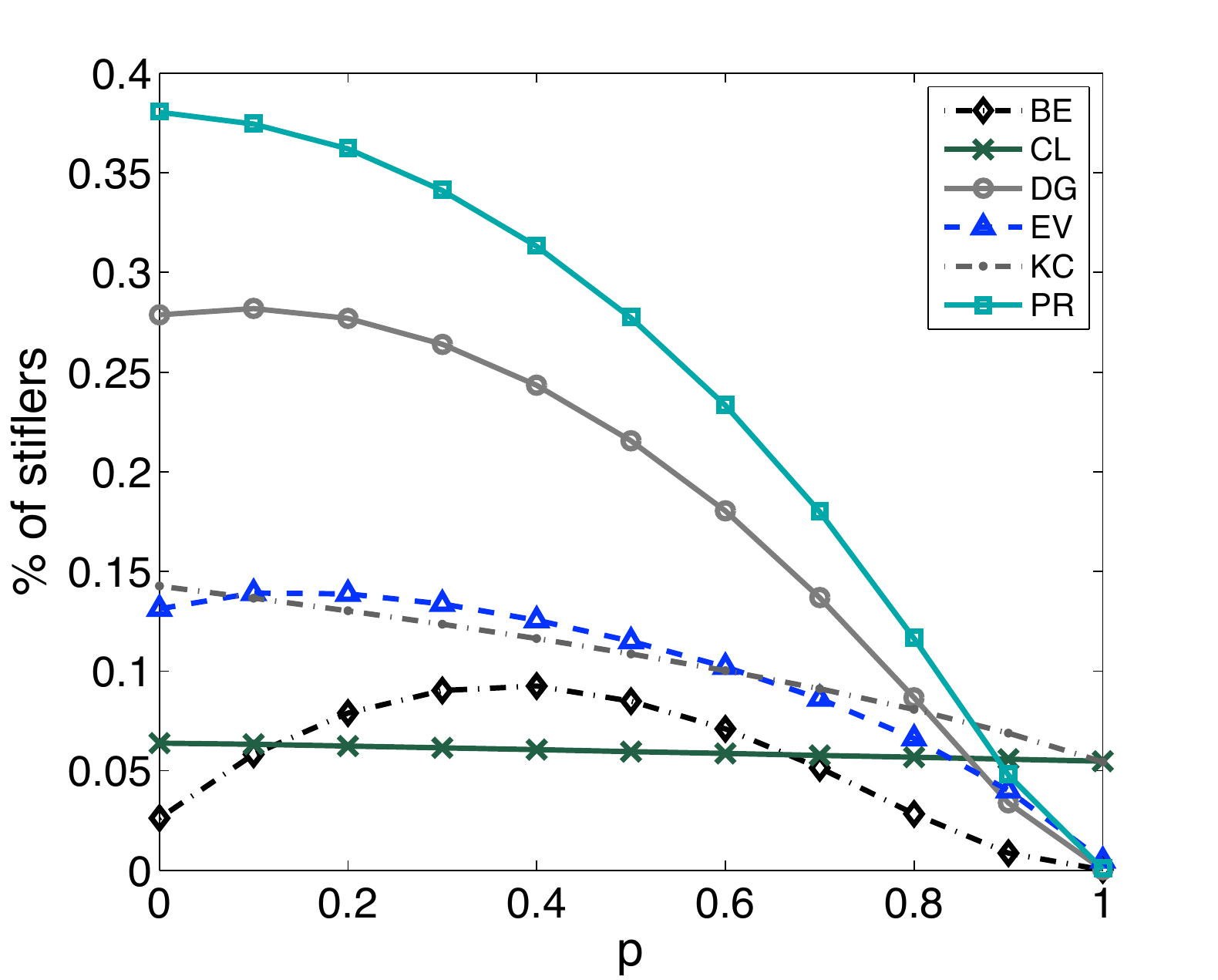}}  
\caption{\label{fig:resCorr} (Color online) Final density of stiflers according to the level of heterogeneity defined by the interpolation function (Eq.~\ref{eq:hetri2}) in social networks: (a) \emph{netscience}, (b) \emph{email}, (c) \emph{Hamsterster} and (d) \emph{Google+}. The centrality measures considered are betweenness centrality (BE), closeness centrality (CL), degree (DG), eigenvector centrality (EV), k-core (KC) and PageRank (PR). The homogeneous propagation is given for $p=1$, whereas the completely heterogeneous case occurs for $p=0$.}
\end{figure*}

We see in Figure~\ref{fig:resCorr} that the curve of the final fraction of stiflers is similar to that verified in scale-free network models (see also Figure \ref{fig:ArtificalCPaper}). For the betweenness centrality and eigenvector centrality, the partial correlation provides the highest fraction of informed individuals. This phenomenon is similar to that obtained for the betweenness centrality in artificial networks (see Figure \ref{fig:Bet-Dg}). On the other hand, for the degree and PageRank, the heterogeneous propagation yields the highest fraction of informed subjects. For the  k-core and closeness centrality, the final fraction of stiflers is not strongly affected by the level of heterogeneity, as verified for the artificial networks. This tendency is due to the close spacing values provided by these measures~\cite{newman2010networks}.

Comparing the centrality measures, we notice that none of them provide the widest reach of rumor propagation in all networks. For the \emph{netscience}, \emph{email} and \emph{Hamster} networks, the partial correlation between the betweenness centrality and the transmission probability provides the highest fraction of stiflers. On the other hand, the correlation with PageRank yields the highest fraction of informed nodes in \emph{Google+} network. These results are different from those observed in the network models, in which the probability of propagation correlated with degree provides the widest reach (see Figure~\ref{fig:ArtificalCPaper}(c) and (d)). This fact is due to the highly structured organization of social networks, that present modular organization, nonvanishing clustering coefficient and degree-degree correlation~\cite{newman2010networks, costa011}. In addition, these networks present peripheral hubs~\cite{Kitsak2010}, contrariwise to Barabási-Albert scale-free networks, in which hubs are central.

\subsection{Controlling rumor spreading}\label{sec:application} 

The study of information diffusion in social networks has practical applications. For instance, in viral marketing, fashion companies promotes new products through its common advertising channels. However, these companies also select a small number of co-participant users from the blogosphere to experience their products for free. Such allied users are expected to enjoy the product and start posting to influence their friends and followers~\cite{Bakshy2012}. Then, these individuals start to propagate information about the companies more often than other users who have not enjoyed the products, improving their propagation capacity. The main issue, in this case, is which role the companies should take on to select the most influential spreaders to improve the promotion and sale of products and services.

Here, we simulate this situation by considering heterogeneous propagation. We start the simulation by setting $\beta_x = 0.5$  and $\mu_x = 0.1$ for all nodes $x=1,\ldots, N$. The rumor spreading start at only one seed node in each simulation. After informing $\eta = $ 1\% of the network, the control technique is applied, simulating the co-participant user's activity. The propagation probability of a set of target nodes (${\tau}$) is increased to $\beta_x = 0.9$. These nodes represent 0.5\% of the population and are selected according to the PageRank measure, from the most central to the most peripheral. The reach of the rumor spreading is measured by the final density of stiflers.

\begin{figure*}[tb]
\centering	
  \subfigure[]{\label{fig:modelAPPa}\includegraphics[width= 0.48\textwidth]{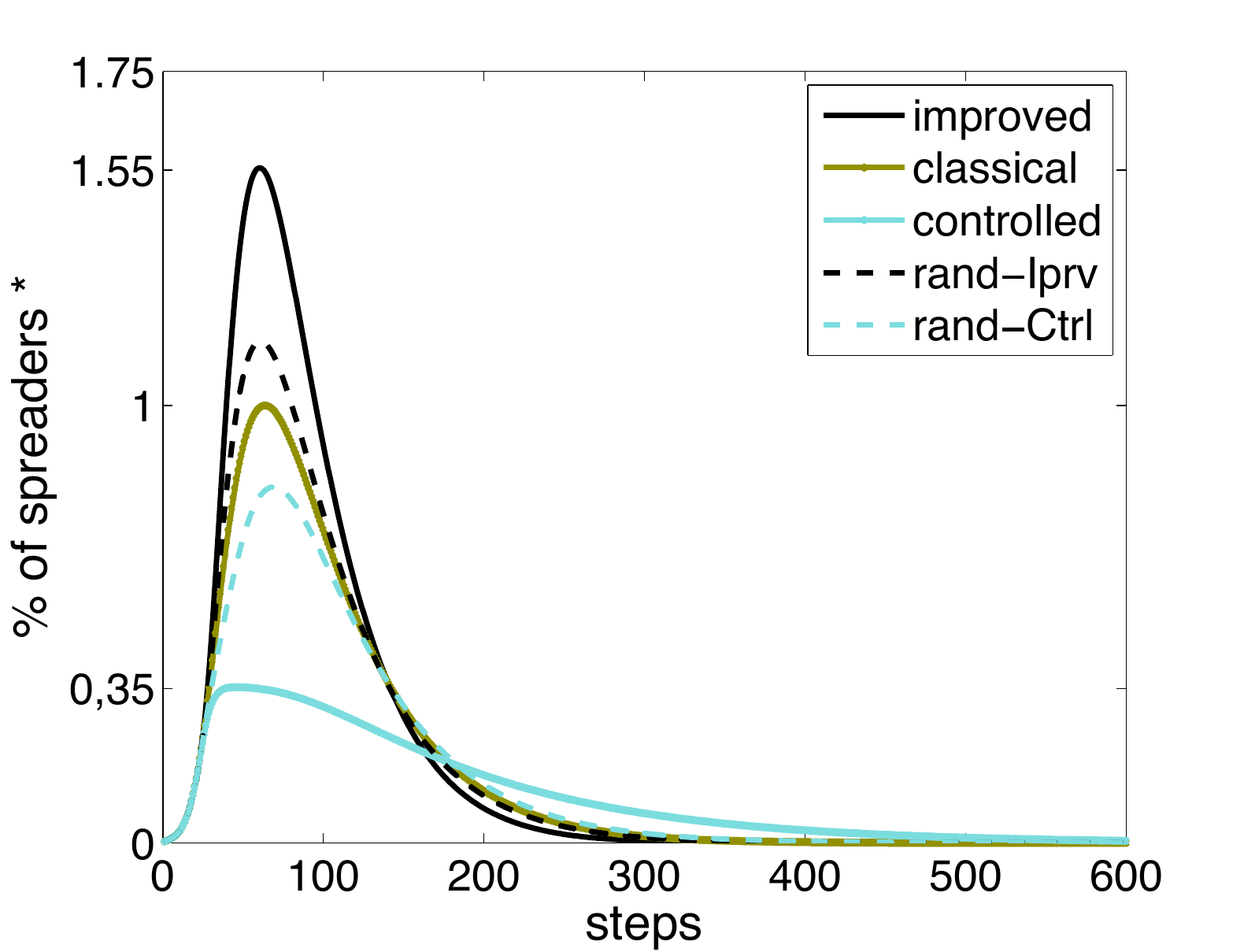}}    
  \subfigure[]{\label{fig:modelAPPb}\includegraphics[width= 0.48\textwidth]{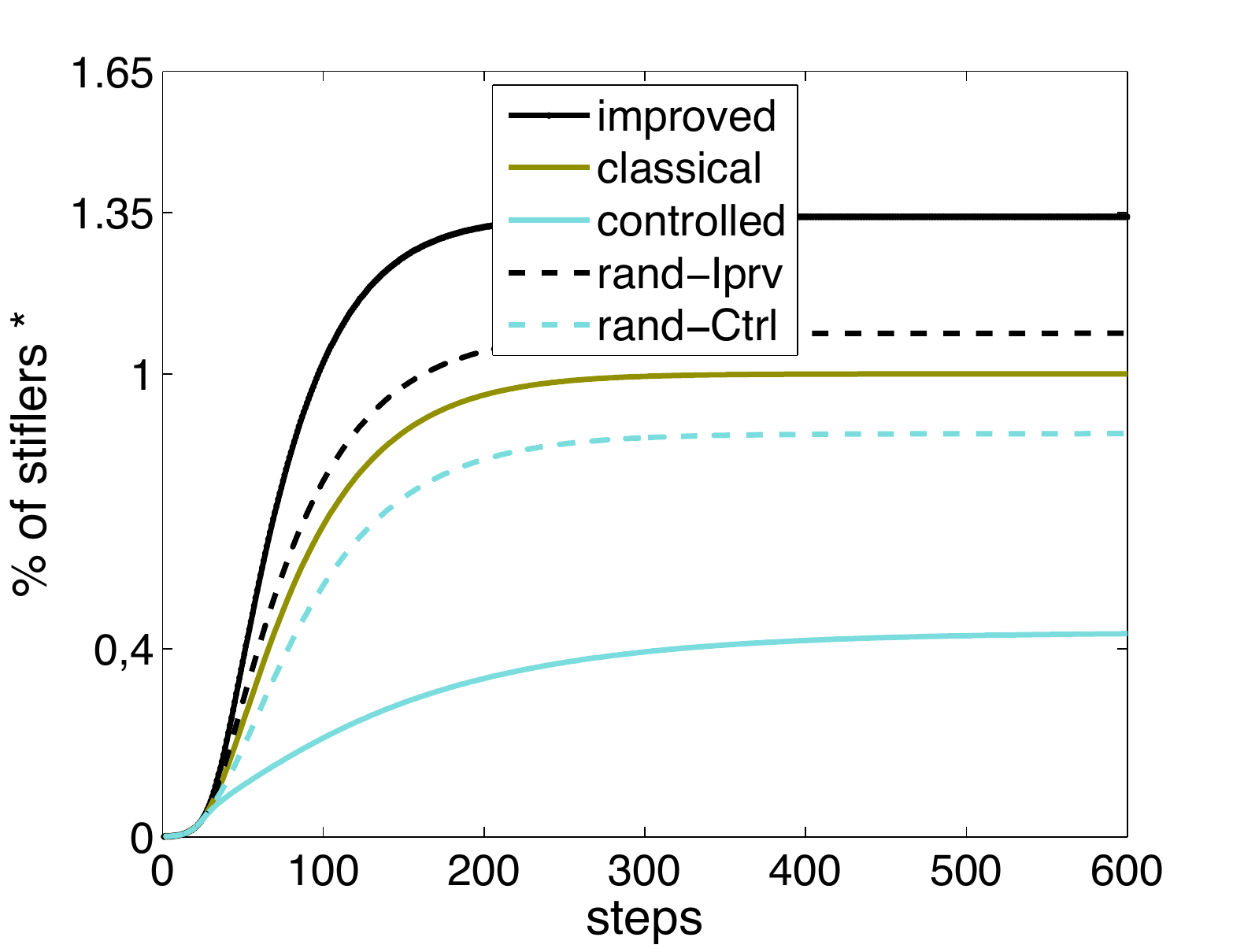}}   
\caption{(Color online) Heterogeneous propagation on the Google+ network. (a) Evolution of the density of spreader normalized by the maximum fraction of spreaders verified in the homogeneous case, and (b) time evolution of the density of informed individuals normalized by the homogeneous spreading. The classical curves shows the results for homogeneous propagation, i.e., $\beta_x = 0.5$ and $\mu_x = 0.1$, $x=1,\ldots, N$. The control strategies are performed by selecting 0.5\% of nodes with the highest PageRank. The selection is also performed by uniform selection of target nodes for improving (rand-Iprv) or decreasing (rand-Ctrl) their propagation probabilities.}
\end{figure*}

Figure~\ref{fig:modelAPPa} shows the evolution of the percentage of spreaders normalized by the maximum fraction of spreaders observed in the homogeneous classical case. Notice that if this percentage is higher than one, then the percentage of informed individuals is higher than that obtained in the classical homogeneous case. The results show that the increase of the capacity of the central nodes makes the rumor propagation faster, reaching a larger number of vertices than the homogeneous case in the same time interval. Indeed, the percentage of spreaders is increased by about 55\% compared to the homogeneous case. We also consider the case in which the target spreaders are selected at random, without considering central properties. In this case, the percentage of spreaders is increased in about 5\%. Figure~\ref{fig:modelAPPb} presents the time evolution of the percentage of informed individuals (stiflers and spreaders). In this case, the heterogeneous propagation based on the central nodes improve this fraction in about 35\%, whereas the random selection only in 5\%. Hence, the selection of the 0.5\% most central nodes enables a very fast and efficient information propagation.

The selection of the central nodes is also performed by considering the betweenness centrality and degree. However, as we see in Table~\ref{tab:spreadingCapacity}, the PageRank is the most suitable measure to selected the most influential spreaders, since the largest fraction of informed individuals is obtained when this metric is chosen. In addition,  Table~\ref{tab:spreadingCapacity} provides the relative improvement of the fraction of stiflers compared with the homogeneous case. The longer the delay to select  the central nodes for improving their transmission probability, the smaller the reach of the rumor. Hence, for an efficient information diffusion, it is necessary to employ the control strategy as soon as the rumor start propagating. 

\begin{table}[!t]
	\centering
	\caption{Improvement and decreasing of rumor propagation. The results show the percentage of the final density of informed nodes ($R_\alpha$, where $\alpha$ is a centrality measure) relative to the homogeneous case. The target central nodes are selected according to the centrality measures betweenness (BE), degree (DG) and PageRank (PR). $\eta$ is the fraction of informed nodes when the control strategy is adopted.}
		\begin{tabular}{c c c c c c c c}
		\hline 
	$\eta$	& \multicolumn{3}{c}{ Improvement \ } & & \multicolumn{3}{c}{ \ Decreasing } \\
		$(\%)$ & $R_{BE}$ & $R_{DG}$ & $R_{PR}$ & & $R_{BE}$ & $R_{DG}$ & $R_{PR}$ \\
		\hline
$1$ &  $16.4\%$  & $21.5\%$ & $\mathbf{22.4\%}$ &  & $20.4\%$  &  $28.7\%$  &  $\mathbf{30.3\%}$ \\
$3$ &  $15.4\%$  & $19.8\%$ & $\mathbf{20.9\%}$ &  & $18.7\%$  &  $25.9\%$  &  $\mathbf{27.5\%}$ \\
$6$ &  $12.5\%$  & $16.0\%$ & $\mathbf{16.8\%}$ &  & $15.0\%$  &  $20.2\%$  &  $\mathbf{21.3\%}$ \\
$10$ &  $5.2\%$  & $7.3\%$ & $\mathbf{8.2\%}$ &  & $6.6\%$  &  $8.5\%$  &  $\mathbf{9.3\%}$ \\
$12$ &  $1.4\%$  & $1.2\%$ & $\mathbf{1.4\%}$ &  & $1.7\%$  &  $2.1\%$  &  $\mathbf{2.3\%}$ \\
$15$ &  $0\%$  & $0\%$ & $0\%$ &  & $0\%$  &  $0\%$  &  $0\%$ \\
 	\hline
		\end{tabular}			
	\label{tab:spreadingCapacity}
\end{table}

Influential spreaders are also important to decrease the rumor propagation. If these users have less interest in the information dissemination than other subjects, the reach of the rumor is affected. This situation occurs in social networks in which influential spreaders are convinced that the information disseminated is not important, which reduces their transmission probability. We simulate this case by starting the rumor propagation with all nodes having the same spreading capacity $\beta_x = 0.5$, $x = 1,\ldots, N$. When the rumor reach $\eta = 1\%$ of the population, the propagation probability of the 0.5\% most central nodes is decreased to $\beta_x = 0.1$. Figure~\ref{fig:modelAPPa} shows that the reach of the rumor is decreased in 65\% when this approach is applied, compared to the homogeneous case. The uniform selection of seed nodes, without considering central properties, decreases the reach of the rumor in less than 10\%. In addition, the final fraction of informed individuals decreases in about 60\% when the capacity of the influential spreaders is reduced (see Figure~\ref{fig:modelAPPb}), whereas the uniform selection of nodes reduces in about 5\%. Hence, influential spreaders can be considered not only for improving, but also for decreasing the rumor spreading.

We also consider a configuration in which the probability of the central nodes is changed to be proportional to their centrality measures. In this case, we select the 0.5\% most central vertices according to a centrality measure and set their probability according to the approach adopted in the last section, i.e., $\beta_x =  \theta_x / \max_{y=1,\ldots, N}(\theta_y)$, such that $0\leq\theta_x\leq 1$, $x=1,2,\ldots, N$ and $\theta_x$ is a centrality measure. Table~\ref{tab:spreadingCapacity} shows the comparison of the PageRank with degree and betweenness centrality measures. As in the improvement approach discussed before, PageRank is the most suitable measure to define the most influential spreaders. In addition, the results show that for an efficient control of the rumor spreading, it is necessary to decrease the probability of propagation of the central nodes in the beginning of the propagation. 

\section{Conclusions}\label{sec:final} 

Several models have been developed to understand the information spreading in complex networks~\cite{Pastor-Satorras2015}. Particularly, rumor is spread through the contact of individuals in social networks~\cite{Zanette2001a}. Current works consider that all individuals have the same capacity of rumor spreading, but in social networks, some people are more influential presenting higher capacity of convincing. In this way, it is more natural to model these dynamics by considering heterogeneous transmission, in which nodes have a spreading capacity given by a defined probability distribution. In this present paper, we assumed that the most central subjects have the highest transmission probability. We provided a model formulation based on the Markov chain formalism and simulated the model in complex networks, including artificial and social networks.

In artificial networks, the complete heterogeneous case, i.e., when the probability of contagion is perfectly correlated with a centrality measure, yields the most efficient propagation. When this probability is proportional to PageRank and degree, it is observed the widest reach of the rumor. In social networks, betweenness centrality and PageRank yield the largest number of informed individuals. Particularly, for the betweenness centrality, a partial correlation between the centrality measure and the transmission probability provides the largest percent of stifler. This effect is due to the presence of ineffective propagators. We also verified that the correlation between the node centrality measures and transmission probability is a necessary condition to improve the rumor propagation.

Finally, we provided a study of rumor dynamics by adjusting the transmission probability of a small number of central nodes. The rumor reach was improved in 35\% when the transmission probability of 0.5\% central nodes is increased. On the other hand, the fraction of informed individuals is decreased in 65\% by reducing the spreading capacity of these nodes. We also verified that for an efficient information propagation, it is necessary to employ the control strategy as soon as the rumor start propagating.

Our results contribute to the understanding of the interplay between rumor propagation and heterogeneous transmission in social networks. The analysis presented here may be extended to other rumor models, such as those with apathy~\cite{Borge-Holthoefer2012a} and lost of interest~\cite{Nekovee2007}. The application in multilayers~\cite{Kivela014, Boccaletti014} and adaptive networks~\cite{Gross08} is also a promising research.

\section*{Appendix}

\subsection*{Network centrality}\label{sec:definition}

Several measures have been developed for network characterization~\cite{costa07}. Particularly, centrality measures aim at ranking the importance of the vertices according to their influence on the network~\cite{newman2003}. However, the definition of centrality is subjective and based on several different network properties~\cite{Arruda014, Morone2015}. For instance, whereas the degree is defined by the number of connections, which is related to local information flow, betweenness centrality considers the load over the whole network~\cite{newman2010networks}. 

Degree centrality (\emph{DG}) is defined by the number of connections of each vertex. The most central nodes are the most connected in the network. The degree of a node $x$ is calculated in terms of the adjacency matrix $\mathbf{A}$,
\begin{equation}\label{eq:k_i}
k_x = \sum_{y=1}^N A_{xy},
\end{equation}
where $A_{xy}= 1$ if $x$ and $y$ are connected, or $A_{xy}= 0$ otherwise. 

The shortest distance between pairs of vertices can also be used to define the level of centrality. Closeness centrality (\emph{CL}) measures the mean distance from a vertex to other ones~\cite{newman2010networks}, i.e., 
\begin{equation}
\label{eq:clsnss}
\displaystyle c_x =  \frac{N}{\sum_{y \neq x} \ell_{xy}} \:, 
\end{equation} 
where $\ell_{xy}$ is the length of the shortest path from $x$ to $y$. 

Centrality is also defined in terms of load, assuming that the information is transmitted through the shortest distances. Betweenness centrality (\emph{BE}) measures this load by the fraction of shortest paths between all pairs of vertices that pass through each node $y$~\cite{freeman77}, i.e.,
\begin{equation}\label{eq:btwns}
 B_y  = \sum_{x,z \in V , x \neq z} \frac{\sigma_{xz}(y)}{\sigma_{xz}} \ ,
\end{equation} 
where $\sigma_{xz}$ is the total number of shortest paths from node $x$ to $z$ and $\sigma_{xz}(y)$ is the number of those paths that pass through $y$.

Eigenvector centrality (\emph{EV}) defines the centrality of a node based on the centrality of its neighbors~\cite{newman2010networks}, i.e., the level of importance of a node is a linear function of the centrality of the vertices to whom its is connected. Mathematically, the centrality is defined by the leading eigenvector of the adjacency matrix $\mathbf{A}$, $\mathbf{A} \mathbf{v} = \alpha \mathbf{v}$, where $\alpha$ is the leading eigenvalue of $\mathbf{A}$ and the element $v_x$ is the eigenvector centrality of node $x$. Thus,  EV is related to principal components analysis (PCA), since it is based on eigenvector decompositions of similarity matrices.

Similar to the eigenvector centrality, PageRank (\emph{PR}) is defined in terms of the spectra of connection-related matriz, i.e., the Google matrix~\cite{BrinPage}. The basic idea is to simulate the user navigation on the World Wide Web. The user follows the links available on the current page or jumps to a new address by typing a new URL, according to a defined probability. Numerically, we consider the power method to calculate the PR, \emph{i.e.}, $ \mathbf{\pi}^t = \mathbf{\pi}^{t-1} \mathbf{G} $, where $\mathbf{\pi}^t$ yields the  PageRank values at step $t$ and  $\mathbf{G} $ is known as the Google  matrix~\cite{BrinPage, newman2010networks}. The PR of a node $x$ for undirected networks is given by
\begin{equation}
\label{eq:pagerank}
{\mathbf{\pi_x}}^t =  \alpha \sum_{y =1}^N \frac{A_{xy}{\pi_y}^{t-1}}{\tilde{k}_y} \ ,
\end{equation} 
where $\tilde{k}_y$ is the degree of its neighbors that attenuates the importance of $x$. When $t = 0$, all vertices have the same PR, $\mathbf{\overrightarrow{\pi}}^0 = [ 1/N, \ldots, 1/N]_{1 \times N}$. $\tilde{k}_y = {k_y} + \delta({k_y}, 0)(1/N)$ is a stochastic adjustment and $\delta(a,b)$ is the Kronecker delta function. Jumps in the navigation occur according to probability $\alpha$. Notice that $\sum_x{\pi_x}^t = 1$ in all iterations. The value $\alpha=0.85$ adopted here is the same as the one defined in the original version of the PR algorithm~\cite{BrinPage}.

$K$-core centrality (\emph{KC}) describes the topology of the network in terms of subnetwork decomposition into cores~\cite{k_core:seidman83}. The core of order $k$ ($H_k$) is the set of vertices whose degree $k_x \geq k$, $x \in H_k$ --- $k$ is the maximum core that $x$ belongs, with $Kc(x) = k$, and $H_k$ is the largest set of vertices with this property~\cite{k_core:seidman83}. The main core is the set of vertices with the largest core order. Such vertices are considered as the most central and play a role as influential spreaders~\cite{Kitsak2010}. Vertices of low values of $k$-core are low connected or at the periphery of the network. Thus, hubs may have low $k$-core values whether they are at the periphery of the network~\cite{Kitsak2010}.

\section{Acknowledgments}
D.A.VO acknowledges CNPq (grant 140688/2013-7). F.A.R. acknowledges CNPq (grant 305940/2010-4) and FAPESP (grant 13/26416-9). L. da F. Costa thanks CNPq (307333/2013-2), NAP-PRP-USP and FAPESP (11/50761-2) for the financial support.

\bibliographystyle{apsrev}
\bibliography{references}

\begin{thebibliography}{44}
\expandafter\ifx\csname natexlab\endcsname\relax\def\natexlab#1{#1}\fi
\expandafter\ifx\csname bibnamefont\endcsname\relax
  \def\bibnamefont#1{#1}\fi
\expandafter\ifx\csname bibfnamefont\endcsname\relax
  \def\bibfnamefont#1{#1}\fi
\expandafter\ifx\csname citenamefont\endcsname\relax
  \def\citenamefont#1{#1}\fi
\expandafter\ifx\csname url\endcsname\relax
  \def\url#1{\texttt{#1}}\fi
\expandafter\ifx\csname urlprefix\endcsname\relax\def\urlprefix{URL }\fi
\providecommand{\bibinfo}[2]{#2}
\providecommand{\eprint}[2][]{\url{#2}}

\bibitem[{\citenamefont{Kosfeld}(2005)}]{Kosfeld05}
\bibinfo{author}{\bibfnamefont{M.}~\bibnamefont{Kosfeld}},
  \bibinfo{journal}{Journal of Mathematical Economics}
  \textbf{\bibinfo{volume}{41}}, \bibinfo{pages}{646} (\bibinfo{year}{2005}).

\bibitem[{\citenamefont{Kimmel}(2004)}]{Kimmel04}
\bibinfo{author}{\bibfnamefont{A.~J.} \bibnamefont{Kimmel}},
  \emph{\bibinfo{title}{Rumors and rumor control: A manager's guide to
  understanding and combatting rumors}} (\bibinfo{publisher}{Routledge},
  \bibinfo{year}{2004}).

\bibitem[{\citenamefont{Kaminsky and Schmukler}(1999)}]{Kaminsky99}
\bibinfo{author}{\bibfnamefont{G.~L.} \bibnamefont{Kaminsky}} \bibnamefont{and}
  \bibinfo{author}{\bibfnamefont{S.~L.} \bibnamefont{Schmukler}},
  \bibinfo{journal}{Journal of International Money and Finance}
  \textbf{\bibinfo{volume}{18}}, \bibinfo{pages}{537} (\bibinfo{year}{1999}).

\bibitem[{\citenamefont{Schindler}(2007)}]{Schindler07}
\bibinfo{author}{\bibfnamefont{M.}~\bibnamefont{Schindler}},
  \emph{\bibinfo{title}{Rumors in financial markets: Insights into behavioral
  finance}}, vol. \bibinfo{volume}{413} (\bibinfo{publisher}{John Wiley \&
  Sons}, \bibinfo{year}{2007}).

\bibitem[{\citenamefont{Pastor-Satorras and
  Vespignani}(2001)}]{Pastor-Satorras2001}
\bibinfo{author}{\bibfnamefont{R.}~\bibnamefont{Pastor-Satorras}}
  \bibnamefont{and}
  \bibinfo{author}{\bibfnamefont{A.}~\bibnamefont{Vespignani}},
  \bibinfo{journal}{Physical Review Letters} \textbf{\bibinfo{volume}{86}},
  \bibinfo{pages}{3200} (\bibinfo{year}{2001}).

\bibitem[{\citenamefont{Hethcote}(2006)}]{Hethcote2006}
\bibinfo{author}{\bibfnamefont{H.~W.} \bibnamefont{Hethcote}},
  \bibinfo{journal}{SIAM Review}  (\bibinfo{year}{2006}).

\bibitem[{\citenamefont{Castellano et~al.}(2009)\citenamefont{Castellano,
  Fortunato, and Loreto}}]{Castellano2009}
\bibinfo{author}{\bibfnamefont{C.}~\bibnamefont{Castellano}},
  \bibinfo{author}{\bibfnamefont{S.}~\bibnamefont{Fortunato}},
  \bibnamefont{and} \bibinfo{author}{\bibfnamefont{V.}~\bibnamefont{Loreto}},
  \bibinfo{journal}{Reviews of Modern Physics} \textbf{\bibinfo{volume}{81}},
  \bibinfo{pages}{591} (\bibinfo{year}{2009}).

\bibitem[{\citenamefont{Funk et~al.}(2010)\citenamefont{Funk, Salath\'{e}, and
  Jansen}}]{Funk2010}
\bibinfo{author}{\bibfnamefont{S.}~\bibnamefont{Funk}},
  \bibinfo{author}{\bibfnamefont{M.}~\bibnamefont{Salath\'{e}}},
  \bibnamefont{and} \bibinfo{author}{\bibfnamefont{V.~A.~A.}
  \bibnamefont{Jansen}}, \bibinfo{journal}{Journal of the Royal Society
  Interface} \textbf{\bibinfo{volume}{7}}, \bibinfo{pages}{1247}
  (\bibinfo{year}{2010}).

\bibitem[{\citenamefont{Pastor-Satorras
  et~al.}(2015)\citenamefont{Pastor-Satorras, Castellano, {Van Mieghem}, and
  Vespignani}}]{Pastor-Satorras2015}
\bibinfo{author}{\bibfnamefont{R.}~\bibnamefont{Pastor-Satorras}},
  \bibinfo{author}{\bibfnamefont{C.}~\bibnamefont{Castellano}},
  \bibinfo{author}{\bibfnamefont{P.}~\bibnamefont{{Van Mieghem}}},
  \bibnamefont{and}
  \bibinfo{author}{\bibfnamefont{A.}~\bibnamefont{Vespignani}},
  \bibinfo{journal}{Reviews of Modern Physics} \textbf{\bibinfo{volume}{87}},
  \bibinfo{pages}{925} (\bibinfo{year}{2015}).

\bibitem[{\citenamefont{Daley and Kendall}(1964)}]{DALEY1964}
\bibinfo{author}{\bibfnamefont{D.~J.} \bibnamefont{Daley}} \bibnamefont{and}
  \bibinfo{author}{\bibfnamefont{D.~G.} \bibnamefont{Kendall}},
  \bibinfo{journal}{Nature} \textbf{\bibinfo{volume}{204}},
  \bibinfo{pages}{1118} (\bibinfo{year}{1964}).

\bibitem[{\citenamefont{Maki and Thompson}(1973)}]{Maki1973}
\bibinfo{author}{\bibfnamefont{D.~P.} \bibnamefont{Maki}} \bibnamefont{and}
  \bibinfo{author}{\bibfnamefont{M.}~\bibnamefont{Thompson}},
  \emph{\bibinfo{title}{Mathematical models and applications: with emphasis on
  the social life, and management sciences}} (\bibinfo{publisher}{Prentice
  Hall}, \bibinfo{year}{1973}).

\bibitem[{\citenamefont{Barab\'{a}si and Albert}(1999)}]{barabasiEalbert1999}
\bibinfo{author}{\bibfnamefont{A.-L.} \bibnamefont{Barab\'{a}si}}
  \bibnamefont{and} \bibinfo{author}{\bibfnamefont{R.}~\bibnamefont{Albert}},
  \bibinfo{journal}{Science} \textbf{\bibinfo{volume}{286}},
  \bibinfo{pages}{509} (\bibinfo{year}{1999}).

\bibitem[{\citenamefont{Newman}(2003)}]{newman2003}
\bibinfo{author}{\bibfnamefont{M.~E.~J.} \bibnamefont{Newman}},
  \bibinfo{journal}{SIAM Review} \textbf{\bibinfo{volume}{45}},
  \bibinfo{pages}{1} (\bibinfo{year}{2003}).

\bibitem[{\citenamefont{Boccaletti et~al.}(2006)\citenamefont{Boccaletti,
  Latora, Moreno, Chavez, and Hwang}}]{boccaletti06}
\bibinfo{author}{\bibfnamefont{S.}~\bibnamefont{Boccaletti}},
  \bibinfo{author}{\bibfnamefont{V.}~\bibnamefont{Latora}},
  \bibinfo{author}{\bibfnamefont{Y.}~\bibnamefont{Moreno}},
  \bibinfo{author}{\bibfnamefont{M.}~\bibnamefont{Chavez}}, \bibnamefont{and}
  \bibinfo{author}{\bibfnamefont{D.-U.} \bibnamefont{Hwang}},
  \bibinfo{journal}{Physics Reports} \textbf{\bibinfo{volume}{424}},
  \bibinfo{pages}{175} (\bibinfo{year}{2006}).

\bibitem[{\citenamefont{Zanette}(2001)}]{Zanette2001a}
\bibinfo{author}{\bibfnamefont{D.~H.} \bibnamefont{Zanette}},
  \bibinfo{journal}{Physical Review E} \textbf{\bibinfo{volume}{64}},
  \bibinfo{pages}{050901} (\bibinfo{year}{2001}).

\bibitem[{\citenamefont{Moreno et~al.}(2004)\citenamefont{Moreno, Nekovee, and
  Pacheco}}]{Moreno2004a}
\bibinfo{author}{\bibfnamefont{Y.}~\bibnamefont{Moreno}},
  \bibinfo{author}{\bibfnamefont{M.}~\bibnamefont{Nekovee}}, \bibnamefont{and}
  \bibinfo{author}{\bibfnamefont{A.~F.} \bibnamefont{Pacheco}},
  \bibinfo{journal}{Physical Review E} \textbf{\bibinfo{volume}{69}},
  \bibinfo{pages}{066130} (\bibinfo{year}{2004}).

\bibitem[{\citenamefont{Nekovee et~al.}(2007)\citenamefont{Nekovee, Moreno,
  Bianconi, and Marsili}}]{Nekovee2007}
\bibinfo{author}{\bibfnamefont{M.}~\bibnamefont{Nekovee}},
  \bibinfo{author}{\bibfnamefont{Y.}~\bibnamefont{Moreno}},
  \bibinfo{author}{\bibfnamefont{G.}~\bibnamefont{Bianconi}}, \bibnamefont{and}
  \bibinfo{author}{\bibfnamefont{M.}~\bibnamefont{Marsili}},
  \bibinfo{journal}{Physica A: Statistical Mechanics and its Applications}
  \textbf{\bibinfo{volume}{374}}, \bibinfo{pages}{457} (\bibinfo{year}{2007}).

\bibitem[{\citenamefont{Meloni et~al.}(2009)\citenamefont{Meloni, Arenas, and
  Moreno}}]{Meloni2009}
\bibinfo{author}{\bibfnamefont{S.}~\bibnamefont{Meloni}},
  \bibinfo{author}{\bibfnamefont{A.}~\bibnamefont{Arenas}}, \bibnamefont{and}
  \bibinfo{author}{\bibfnamefont{Y.}~\bibnamefont{Moreno}},
  \bibinfo{journal}{Proceedings of the National Academy of Sciences of the
  United States of America} \textbf{\bibinfo{volume}{106}},
  \bibinfo{pages}{16897} (\bibinfo{year}{2009}).

\bibitem[{\citenamefont{Jennings}(1937)}]{Jennings37}
\bibinfo{author}{\bibfnamefont{H.}~\bibnamefont{Jennings}},
  \bibinfo{journal}{Sociometry} \textbf{\bibinfo{volume}{1}},
  \bibinfo{pages}{99} (\bibinfo{year}{1937}).

\bibitem[{\citenamefont{Kitsak et~al.}(2010)\citenamefont{Kitsak, Gallos,
  Havlin, Liljeros, Muchnik, Stanley, and Makse}}]{Kitsak2010}
\bibinfo{author}{\bibfnamefont{M.}~\bibnamefont{Kitsak}},
  \bibinfo{author}{\bibfnamefont{L.~K.} \bibnamefont{Gallos}},
  \bibinfo{author}{\bibfnamefont{S.}~\bibnamefont{Havlin}},
  \bibinfo{author}{\bibfnamefont{F.}~\bibnamefont{Liljeros}},
  \bibinfo{author}{\bibfnamefont{L.}~\bibnamefont{Muchnik}},
  \bibinfo{author}{\bibfnamefont{H.~E.} \bibnamefont{Stanley}},
  \bibnamefont{and} \bibinfo{author}{\bibfnamefont{H.~A.} \bibnamefont{Makse}},
  \bibinfo{journal}{Nature Physics} \textbf{\bibinfo{volume}{6}},
  \bibinfo{pages}{888} (\bibinfo{year}{2010}).

\bibitem[{\citenamefont{Pei and Makse}(2013)}]{Pei013}
\bibinfo{author}{\bibfnamefont{S.}~\bibnamefont{Pei}} \bibnamefont{and}
  \bibinfo{author}{\bibfnamefont{H.~A.} \bibnamefont{Makse}},
  \bibinfo{journal}{Journal of Statistical Mechanics: Theory and Experiment}
  \textbf{\bibinfo{volume}{2013}}, \bibinfo{pages}{P12002}
  (\bibinfo{year}{2013}).

\bibitem[{\citenamefont{de~Arruda et~al.}(2014)\citenamefont{de~Arruda,
  Barbieri, Rodr{\'\i}guez, Rodrigues, Moreno, and
  da~Fontoura~Costa}}]{Arruda014}
\bibinfo{author}{\bibfnamefont{G.~F.} \bibnamefont{de~Arruda}},
  \bibinfo{author}{\bibfnamefont{A.~L.} \bibnamefont{Barbieri}},
  \bibinfo{author}{\bibfnamefont{P.~M.} \bibnamefont{Rodr{\'\i}guez}},
  \bibinfo{author}{\bibfnamefont{F.~A.} \bibnamefont{Rodrigues}},
  \bibinfo{author}{\bibfnamefont{Y.}~\bibnamefont{Moreno}}, \bibnamefont{and}
  \bibinfo{author}{\bibfnamefont{L.}~\bibnamefont{da~Fontoura~Costa}},
  \bibinfo{journal}{Physical Review E} \textbf{\bibinfo{volume}{90}},
  \bibinfo{pages}{032812} (\bibinfo{year}{2014}).

\bibitem[{\citenamefont{Pei et~al.}(2014)\citenamefont{Pei, Muchnik,
  Andrade~Jr, Zheng, and Makse}}]{Pei014}
\bibinfo{author}{\bibfnamefont{S.}~\bibnamefont{Pei}},
  \bibinfo{author}{\bibfnamefont{L.}~\bibnamefont{Muchnik}},
  \bibinfo{author}{\bibfnamefont{J.~S.} \bibnamefont{Andrade~Jr}},
  \bibinfo{author}{\bibfnamefont{Z.}~\bibnamefont{Zheng}}, \bibnamefont{and}
  \bibinfo{author}{\bibfnamefont{H.~A.} \bibnamefont{Makse}},
  \bibinfo{journal}{Scientific Reports} \textbf{\bibinfo{volume}{4}}
  (\bibinfo{year}{2014}).

\bibitem[{\citenamefont{Morone and Makse}(2015)}]{Morone2015}
\bibinfo{author}{\bibfnamefont{F.}~\bibnamefont{Morone}} \bibnamefont{and}
  \bibinfo{author}{\bibfnamefont{H.~A.} \bibnamefont{Makse}},
  \bibinfo{journal}{Nature}  (\bibinfo{year}{2015}).

\bibitem[{\citenamefont{Keeling and Eames}(2005)}]{Keeling2005}
\bibinfo{author}{\bibfnamefont{M.~J.} \bibnamefont{Keeling}} \bibnamefont{and}
  \bibinfo{author}{\bibfnamefont{K.~T.~D.} \bibnamefont{Eames}},
  \bibinfo{journal}{Journal of the Royal Society Interface}
  \textbf{\bibinfo{volume}{2}}, \bibinfo{pages}{295} (\bibinfo{year}{2005}).

\bibitem[{\citenamefont{G{\'o}mez-Gardenes
  et~al.}(2011)\citenamefont{G{\'o}mez-Gardenes, G{\'o}mez, Arenas, and
  Moreno}}]{Gomez011}
\bibinfo{author}{\bibfnamefont{J.}~\bibnamefont{G{\'o}mez-Gardenes}},
  \bibinfo{author}{\bibfnamefont{S.}~\bibnamefont{G{\'o}mez}},
  \bibinfo{author}{\bibfnamefont{A.}~\bibnamefont{Arenas}}, \bibnamefont{and}
  \bibinfo{author}{\bibfnamefont{Y.}~\bibnamefont{Moreno}},
  \bibinfo{journal}{Physical Review Letters} \textbf{\bibinfo{volume}{106}},
  \bibinfo{pages}{128701} (\bibinfo{year}{2011}).

\bibitem[{\citenamefont{Peron and Rodrigues}(2012)}]{Peron012}
\bibinfo{author}{\bibfnamefont{T.~K.~D.} \bibnamefont{Peron}} \bibnamefont{and}
  \bibinfo{author}{\bibfnamefont{F.~A.} \bibnamefont{Rodrigues}},
  \bibinfo{journal}{Physical Review E} \textbf{\bibinfo{volume}{86}},
  \bibinfo{pages}{056108} (\bibinfo{year}{2012}).

\bibitem[{\citenamefont{Erd{\"o}s and R{\'e}nyi}(1959)}]{erdosrenyi1959}
\bibinfo{author}{\bibfnamefont{P.}~\bibnamefont{Erd{\"o}s}} \bibnamefont{and}
  \bibinfo{author}{\bibfnamefont{A.}~\bibnamefont{R{\'e}nyi}},
  \bibinfo{journal}{Publicationes Mathematicae} \textbf{\bibinfo{volume}{6}},
  \bibinfo{pages}{290} (\bibinfo{year}{1959}).

\bibitem[{\citenamefont{Watts and Strogatz}(1998)}]{watts98}
\bibinfo{author}{\bibfnamefont{D.~J.} \bibnamefont{Watts}} \bibnamefont{and}
  \bibinfo{author}{\bibfnamefont{S.~H.} \bibnamefont{Strogatz}},
  \bibinfo{journal}{Nature} \textbf{\bibinfo{volume}{393}},
  \bibinfo{pages}{440} (\bibinfo{year}{1998}).

\bibitem[{\citenamefont{Barth\'{e}lemy}(2003)}]{Barthelemy2003}
\bibinfo{author}{\bibfnamefont{M.}~\bibnamefont{Barth\'{e}lemy}},
  \bibinfo{journal}{Europhysics Letters (EPL)} \textbf{\bibinfo{volume}{63}},
  \bibinfo{pages}{915} (\bibinfo{year}{2003}).

\bibitem[{\citenamefont{Costa et~al.}(2007)\citenamefont{Costa, Rodrigues,
  Travieso, and Villas~Boas}}]{costa07}
\bibinfo{author}{\bibfnamefont{L.~d.~F.} \bibnamefont{Costa}},
  \bibinfo{author}{\bibfnamefont{F.~A.} \bibnamefont{Rodrigues}},
  \bibinfo{author}{\bibfnamefont{G.}~\bibnamefont{Travieso}}, \bibnamefont{and}
  \bibinfo{author}{\bibfnamefont{P.~R.} \bibnamefont{Villas~Boas}},
  \bibinfo{journal}{Advances in Physics} \textbf{\bibinfo{volume}{56}},
  \bibinfo{pages}{167} (\bibinfo{year}{2007}).

\bibitem[{\citenamefont{Guimera et~al.}(2003)\citenamefont{Guimera, Danon,
  Diaz-Guilera, Giralt, and Arenas}}]{Guimera03}
\bibinfo{author}{\bibfnamefont{R.}~\bibnamefont{Guimera}},
  \bibinfo{author}{\bibfnamefont{L.}~\bibnamefont{Danon}},
  \bibinfo{author}{\bibfnamefont{A.}~\bibnamefont{Diaz-Guilera}},
  \bibinfo{author}{\bibfnamefont{F.}~\bibnamefont{Giralt}}, \bibnamefont{and}
  \bibinfo{author}{\bibfnamefont{A.}~\bibnamefont{Arenas}},
  \bibinfo{journal}{Physical Review E} \textbf{\bibinfo{volume}{68}},
  \bibinfo{pages}{065103} (\bibinfo{year}{2003}).

\bibitem[{kon(2014)}]{konectURL}
\emph{\bibinfo{title}{{Hamsterster friendships network dataset -- KONECT}}}
  (\bibinfo{year}{2014}), \urlprefix\url{http://konect.uni-koblenz.de}.

\bibitem[{\citenamefont{McAuley and Leskovec}(2012)}]{GplusPaper}
\bibinfo{author}{\bibfnamefont{J.}~\bibnamefont{McAuley}} \bibnamefont{and}
  \bibinfo{author}{\bibfnamefont{J.}~\bibnamefont{Leskovec}}, in
  \emph{\bibinfo{booktitle}{Advances in Neural Information Processing Systems}}
  (\bibinfo{year}{2012}), pp. \bibinfo{pages}{548--556}.

\bibitem[{\citenamefont{Newman}(2010)}]{newman2010networks}
\bibinfo{author}{\bibfnamefont{M.}~\bibnamefont{Newman}},
  \emph{\bibinfo{title}{{Networks: an introduction}}}
  (\bibinfo{publisher}{Oxford University Press}, \bibinfo{year}{2010}).

\bibitem[{\citenamefont{Costa et~al.}(2011)\citenamefont{Costa, Oliveira~Jr,
  Travieso, Rodrigues, Villas~Boas, Antiqueira, Viana, and
  Correa~Rocha}}]{costa011}
\bibinfo{author}{\bibfnamefont{L.~d.~F.} \bibnamefont{Costa}},
  \bibinfo{author}{\bibfnamefont{O.~N.} \bibnamefont{Oliveira~Jr}},
  \bibinfo{author}{\bibfnamefont{G.}~\bibnamefont{Travieso}},
  \bibinfo{author}{\bibfnamefont{F.~A.} \bibnamefont{Rodrigues}},
  \bibinfo{author}{\bibfnamefont{P.~R.} \bibnamefont{Villas~Boas}},
  \bibinfo{author}{\bibfnamefont{L.}~\bibnamefont{Antiqueira}},
  \bibinfo{author}{\bibfnamefont{M.~P.} \bibnamefont{Viana}}, \bibnamefont{and}
  \bibinfo{author}{\bibfnamefont{L.~E.} \bibnamefont{Correa~Rocha}},
  \bibinfo{journal}{Advances in Physics} \textbf{\bibinfo{volume}{60}},
  \bibinfo{pages}{329} (\bibinfo{year}{2011}).

\bibitem[{\citenamefont{Bakshy et~al.}(2012)\citenamefont{Bakshy, Rosenn,
  Marlow, and Adamic}}]{Bakshy2012}
\bibinfo{author}{\bibfnamefont{E.}~\bibnamefont{Bakshy}},
  \bibinfo{author}{\bibfnamefont{I.}~\bibnamefont{Rosenn}},
  \bibinfo{author}{\bibfnamefont{C.}~\bibnamefont{Marlow}}, \bibnamefont{and}
  \bibinfo{author}{\bibfnamefont{L.}~\bibnamefont{Adamic}}, in
  \emph{\bibinfo{booktitle}{Proceedings of the 21st International Conference on
  World Wide Web}} (\bibinfo{year}{2012}), WWW '12, pp.
  \bibinfo{pages}{519--528}.

\bibitem[{\citenamefont{Borge-Holthoefer
  et~al.}(2012)\citenamefont{Borge-Holthoefer, Meloni, Gon\c{c}alves, and
  Moreno}}]{Borge-Holthoefer2012a}
\bibinfo{author}{\bibfnamefont{J.}~\bibnamefont{Borge-Holthoefer}},
  \bibinfo{author}{\bibfnamefont{S.}~\bibnamefont{Meloni}},
  \bibinfo{author}{\bibfnamefont{B.}~\bibnamefont{Gon\c{c}alves}},
  \bibnamefont{and} \bibinfo{author}{\bibfnamefont{Y.}~\bibnamefont{Moreno}},
  \bibinfo{journal}{Journal of Statistical Physics}
  \textbf{\bibinfo{volume}{151}}, \bibinfo{pages}{383} (\bibinfo{year}{2012}).

\bibitem[{\citenamefont{Kivel{\"a} et~al.}(2014)\citenamefont{Kivel{\"a},
  Arenas, Barthelemy, Gleeson, Moreno, and Porter}}]{Kivela014}
\bibinfo{author}{\bibfnamefont{M.}~\bibnamefont{Kivel{\"a}}},
  \bibinfo{author}{\bibfnamefont{A.}~\bibnamefont{Arenas}},
  \bibinfo{author}{\bibfnamefont{M.}~\bibnamefont{Barthelemy}},
  \bibinfo{author}{\bibfnamefont{J.~P.} \bibnamefont{Gleeson}},
  \bibinfo{author}{\bibfnamefont{Y.}~\bibnamefont{Moreno}}, \bibnamefont{and}
  \bibinfo{author}{\bibfnamefont{M.~A.} \bibnamefont{Porter}},
  \bibinfo{journal}{Journal of Complex Networks} \textbf{\bibinfo{volume}{2}},
  \bibinfo{pages}{203} (\bibinfo{year}{2014}).

\bibitem[{\citenamefont{Boccaletti et~al.}(2014)\citenamefont{Boccaletti,
  Bianconi, Criado, Del~Genio, G{\'o}mez-Garde{\~n}es, Romance,
  Sendi{\~n}a-Nadal, Wang, and Zanin}}]{Boccaletti014}
\bibinfo{author}{\bibfnamefont{S.}~\bibnamefont{Boccaletti}},
  \bibinfo{author}{\bibfnamefont{G.}~\bibnamefont{Bianconi}},
  \bibinfo{author}{\bibfnamefont{R.}~\bibnamefont{Criado}},
  \bibinfo{author}{\bibfnamefont{C.~I.} \bibnamefont{Del~Genio}},
  \bibinfo{author}{\bibfnamefont{J.}~\bibnamefont{G{\'o}mez-Garde{\~n}es}},
  \bibinfo{author}{\bibfnamefont{M.}~\bibnamefont{Romance}},
  \bibinfo{author}{\bibfnamefont{I.}~\bibnamefont{Sendi{\~n}a-Nadal}},
  \bibinfo{author}{\bibfnamefont{Z.}~\bibnamefont{Wang}}, \bibnamefont{and}
  \bibinfo{author}{\bibfnamefont{M.}~\bibnamefont{Zanin}},
  \bibinfo{journal}{Physics Reports} \textbf{\bibinfo{volume}{544}},
  \bibinfo{pages}{1} (\bibinfo{year}{2014}).

\bibitem[{\citenamefont{Gross and Blasius}(2008)}]{Gross08}
\bibinfo{author}{\bibfnamefont{T.}~\bibnamefont{Gross}} \bibnamefont{and}
  \bibinfo{author}{\bibfnamefont{B.}~\bibnamefont{Blasius}},
  \bibinfo{journal}{Journal of the Royal Society Interface}
  \textbf{\bibinfo{volume}{5}}, \bibinfo{pages}{259} (\bibinfo{year}{2008}).

\bibitem[{\citenamefont{Freeman}(1977)}]{freeman77}
\bibinfo{author}{\bibfnamefont{L.~C.} \bibnamefont{Freeman}},
  \bibinfo{journal}{Sociometry} \textbf{\bibinfo{volume}{40}},
  \bibinfo{pages}{35} (\bibinfo{year}{1977}).

\bibitem[{\citenamefont{Brin and Page}(1998)}]{BrinPage}
\bibinfo{author}{\bibfnamefont{S.}~\bibnamefont{Brin}} \bibnamefont{and}
  \bibinfo{author}{\bibfnamefont{L.}~\bibnamefont{Page}}, in
  \emph{\bibinfo{booktitle}{Seventh International World-Wide Web Conference
  (WWW 1998)}} (\bibinfo{year}{1998}).

\bibitem[{\citenamefont{Seidman}(1983)}]{k_core:seidman83}
\bibinfo{author}{\bibfnamefont{S.}~\bibnamefont{Seidman}},
  \bibinfo{journal}{Social networks} \textbf{\bibinfo{volume}{5}},
  \bibinfo{pages}{269} (\bibinfo{year}{1983}).

\end{thebibliography}

\end{document}